%% file: main.tex
\documentclass[lettersize,journal]{IEEEtran}
\usepackage{amsmath,amsfonts}
\usepackage{array}
\usepackage[caption=false,font=normalsize,labelfont=sf,textfont=sf]{subfig}
\usepackage{textcomp}
\usepackage{stfloats}
\usepackage{verbatim}
\usepackage{graphicx}
\usepackage{cite}

\usepackage{picinpar}
\usepackage{amsmath}
\usepackage[hyphens]{url}
\usepackage{flushend}
\usepackage[utf8]{inputenc}
\usepackage{colortbl}
\usepackage{soul}
\usepackage{multirow}
\usepackage{pifont}
\usepackage{xcolor}
\usepackage{alltt}
\usepackage[hidelinks]{hyperref}
\usepackage{enumerate}
\usepackage{siunitx}
\usepackage{breakurl}
\usepackage{epstopdf}
\usepackage{pbox}
\usepackage[acronym]{glossaries}
\usepackage{hyperref}
\usepackage[capitalise, noabbrev]{cleveref}
\usepackage{makecell}
\usepackage{amssymb}
\usepackage{pifont}
\usepackage{threeparttable}
\newcommand{\cmark}{\ding{51}}%
\newcommand{\xmark}{\ding{55}}%
\usepackage{refcount}
\usepackage[htt]{hyphenat}
\usepackage{listings}
\usepackage[pscoord]{eso-pic}
\usepackage{orcidlink}

\begin{document}

\title{CV32RT: Enabling Fast Interrupt and Context Switching for RISC-V Microcontrollers}

\author{Robert~Balas\orcidlink{0000-0002-7231-9315},~\IEEEmembership{Student Member, IEEE},
        Alessandro~Ottaviano\orcidlink{0009-0000-9924-3536},~\IEEEmembership{Student Member, IEEE},
        and~Luca~Benini\orcidlink{0000-0001-8068-3806},~\IEEEmembership{Fellow, IEEE}
        \IEEEcompsocitemizethanks{\IEEEcompsocthanksitem R.~Balas, A.~Ottaviano, and L.~Benini are with the Integrated Systems Laboratory (IIS), ETH Zurich, Switzerland.\protect\\
        E-mail: \{balasr, aottaviano,lbenini\}@iis.ee.ethz.ch
        \IEEEcompsocthanksitem L.~Benini is also with the Department of Electrical, Electronic and Information Engineering (DEI), University of Bologna, Bologna, Italy.\protect
        \IEEEcompsocthanksitem This work was supported in part through the TRISTAN (101095947) project that received funding from the HORIZON KDT-JU programme.\protect
        }
}



\maketitle
    \AddToShipoutPictureFG*{%
        \put(%
            \paperwidth-6mm,%
            \paperheight-1cm%
            ){\vtop{{\null}\makebox[0pt][c]{%
                \rotatebox[origin=c]{90}{%
                    This work has been submitted to the IEEE for possible publication. Copyright may be transferred without notice, after which this version may no longer be accessible.%
                }%
            }}%
        }%
    }


\begin{abstract}
Processors using the open RISC-V ISA are finding increasing adoption in the embedded world. Many embedded use cases have real-time constraints and require flexible, predictable, and fast reactive handling of incoming events. However, RISC-V processors are still lagging in this area compared to more mature proprietary architectures, such as ARM Cortex-M and TriCore, which have been tuned for years. The default interrupt controller standardized by RISC-V, the Core Local Interruptor (CLINT), lacks configurability in prioritization and preemption of interrupts. The RISC-V Core Local Interrupt Controller (CLIC) specification addresses this concern by enabling preemptible, low-latency vectored interrupts while also envisioning optional extensions to improve interrupt latency. In this work, we implement a CLIC for the CV32E40P, an industrially supported open-source 32-bit MCU-class RISC-V core, and enhance it with \texttt{fastirq}: a custom extension that provides interrupt latency as low as 6 cycles. We call CV32RT our enhanced core. To the best of our knowledge, CV32RT is the first fully open-source RV32 core with competitive interrupt-handling features compared to the Arm Cortex-M series and TriCore. 
The proposed extensions are also demonstrated to improve task context switching in real-time operating systems.

\end{abstract}

\input{acronyms.tex}


\section{Introduction}

\IEEEPARstart{S}{everal} markets, from automotive to aerospace and robotics, rely on real-time hardware and software~\cite{rochange_2014}. 
For example, the automotive industry employs hundreds of \glspl{ecu} for real-time applications, such as electronic engine control, gearbox control, cruise control, anti-lock brake systems, and many other tasks.

\Glspl{gpos} are typically tuned for average throughput rather than real-time requirements imposed by such scenarios. For example, the development of Linux, a popular open-source \gls{gpos} kernel, is focused on average performance, making it less suitable to be used for real-time applications~\cite{marongiu_survey_2018}.

Albeit extensions and modifications which aim at improving determinism and latencies of critical operations in Linux have been proposed and implemented~\cite{linux-preempt-rt, miao_2011, marongiu_survey_2018, rtai_2000}, they do not guarantee strict bounds on maximum latencies of operations and lack industry-grade maturity to be employed in hard real-time scenarios.

\Glspl{rtos} kernels are special-purpose \glspl{os} designed to provide real-time guarantees, such as task scheduling according to a given expected completion deadline and deterministic latencies of various operations~\cite{ramam_94}.

The RTOS' scheduler might add a significant overhead due to the combined effect of both the \textit{context switches} required to handle the transition from a foreground to a background task and the amount of time elapsed from the source event that causes the preemption and the first instruction of the awakened task (known as \textit{interrupt latency}~\cite{minimize_int}), thus increasing the \gls{wcet}~\cite{intro_embedded, marongiu_survey_2018, WCET_REALLOCATION_TVLSI}.

The cost of saving and restoring the task state during a context switch is a significant concern as it remains relatively high. For instance, the process state that needs to be saved on a context switch includes the program counter, register files, status registers, address space mapping, etc. 
Therefore, a significant number of memory access operations need to be performed to \textit{store} the state of the preempted task and \textit{restore} the state of
the new task to be executed~\cite{xiangrong_2006}.
Long context switch times reduce available \textit{task utilization} and the minimum viable switching granularity.

Besides the task's context switching, a switch into an interrupt context from normal program execution happens each time an asynchronous event is triggered from, e.g., an I/O peripheral device.
Behnke et al. \cite{behnke_2020} provide a meaningful example of the impact on \textit{interrupt latency} of network loads in time-critical \glspl{mcu}, where a high overhead generated by the receiving of packets can be seen in continuous floods as well as short transmission bursts, reaching up to 50\% task \textit{lateness} --- the additional time a task takes to finish than its deadline allows, which will start to accumulate over iterations from the critical network load --- increase per packet per second.
Furthermore, hardware-induced interrupt latency is only one part of the problem: software-induced interrupt latency, introduced by the \gls{gpos}/\gls{rtos} scheduler and the user code, primarily impacts the capability of the system to provide timely responses to asynchronous events.

Low interrupt latency and context switch time are crucial metrics for a wide range of platforms ranging from commodity \gls{mcu}-class embedded systems to more advanced and complex application-class \glspl{mcs}, where time/safety-critical and non-critical applications coexist on different isolated partitions of the same hardware platform~\cite{AUTOMOTIVE_DYNAMIQ_MPAM}.

\textit{Response} and \textit{context switch} time minimization thus become a challenge to be tackled at the \gls{hw} / \gls{sw} interface, where \gls{sw} programming techniques and \gls{hw} \textit{interrupt controller} architectures can cooperate to ensure minimal response time~\cite{minimize_int}.

Although commercial vendors and IP providers offer such features as in-house solutions, they are often proprietary and tightly coupled with the vendor's \gls{isa}, target hardware family, and associated software stack.
On the other hand, in the last decade, the ever-growing RISC-V ecosystem~\cite{asanovic_riscv} has been offering a modular, free, and open-source \gls{isa} which is rapidly becoming the \textit{de facto} \textit{lingua franca} of computing.

Nevertheless, RISC-V support for fast interrupt and context switch handling is still not mature enough to compete with incumbent proprietary architectures, as it lacks flexible interrupt prioritization, preemption mechanisms, and low interrupt latency. 
For this reason, the RISC-V community has been developing an extension to the \textit{Privileged} specifications~\cite{RISCV_II} with the proposal of the RISC-V \gls{clic}~\cite{clic}, currently under ratification by the community, to handle such real-time scenarios.

While the modular RISC-V \gls{isa} enables developing orthogonal \textit{custom extensions}, to the best of the authors' knowledge, there are few published works that (i) tackle the problem of minimizing \textit{interrupt latency} and \textit{context switch} time from a holistic (\gls{hw} and \gls{sw}) viewpoint for RISC-V, (ii) try to close the gap with more established proprietary solutions, thereby promoting RISC-V as a valuable candidate for time- and safety-critical application domains such as automotive and aerospace, and (iii) provide an open-source solution to be shared with the community.

To address these open challenges, we propose CV32RT, a 32-bit RISC-V core that extends the interrupt handling capabilities of CV32E40P~\cite{RISCY_CORE_TVLSI, CV32E40P_manual}, an industrially supported open-source core, to achieve best-in-class interrupt latency and fast context switching against \gls{cots} processor vendors, paving the road for RISC-V architectures in time-critical systems.

In particular, this paper makes the following contributions:

\begin{itemize}
    \item We replace the CV32E40P interrupt controller with an implementation of the RISC-V \gls{clic} specification, which provides the core with key features such as prioritization by level and priority, selective hardware vectoring, and non-nested interrupt optimization (\textit{tail-chaining} through the \texttt{xnxti}~\cite{clic} CSR) directly as RISC-V standard extension (\cref{subsec:cv32rt_archi_overview}).
    We make the implementation available under a permissive open-source license~\footnote{\url{https://github.com/pulp-platform/clic}}.

    \item We design a \textit{fast interrupt extension} (\texttt{fastirq}) to accelerate both nested (\cref{subsec:cv32rt_archi_overview}) and non-nested (\cref{subsec:int_scenarios}) interrupt case scenarios.
    \texttt{fastirq} reduces \textit{interrupt latency} by hiding the latency through memory banks and a background saving mechanism. The same mechanism allows one also to accelerate \textit{context switching} through \gls{hw}/\gls{sw} cooperation.
    Furthermore, we propose \texttt{early mret} (\texttt{emret}), a novel instruction that further optimizes tail-chaining compared to the baseline strategy proposed in the \gls{clic} standard (i.e., \texttt{xnxti}) and its enhancement from~\cite{nuclei_isa} (\texttt{jalxnxti}).
    
    \item We integrate CV32RT within an open-source \gls{soc}~\cite{Ottaviano2023ControlPULPAR}~\footnote{\url{https://github.com/pulp-platform/control-pulp}}. We evaluate CV32RT interrupt handling capabilities while showing the negligible area overhead introduced by the extension in a modern technology node against its performance benefits (\cref{sec:evaluation}).

    \item Finally, we compare CV32RT with leading \gls{cots} systems in both nested and non-nested interrupt scenarios (\cref{sec:related_works}). We show that the proposed solution promotes RISC-V as a competitive candidate for building the next generation of time-critical systems.

\end{itemize}

\section{Background}\label{sec:background}

We give a brief overview of the full-system platform used to design and implement the proposed interrupt extension in \cref{subsec:platform}, motivate and explain the relevant target metrics in \cref{subsec:latency}, and describe the current status of interrupt handling in RISC-V in \cref{subsec:rv32-int}.

\subsection{RISC-V system platform}\label{subsec:platform}
We rely on the CV32E40P core~\cite{RISCY_CORE_TVLSI} --- abbreviated to CV32 in this paper --- an open-source, industry-grade, 32-bit, in-order, four-stage RISC-V core, as the basis for implementing our extensions. This core is embedded in ControlPULP~\cite{Ottaviano2023ControlPULPAR}, a \gls{soc} specialized in running real-time workloads. The system contains a CV32 manager core, a programmable accelerator subsystem consisting of 8 CV32 cores, and a set of standard peripherals such as QSPI, I2C, and UART.
The manager core is responsible for scheduling tasks, communicating with the peripherals, offloading tasks to the accelerator subsystem, and being responsive to asynchronous external events, e.g., interrupts.

ControlPULP hosts a set of \glspl{spm} that guarantee \textit{single-cycle access time} from the CV32 manager core.
This design choice enables deterministic memory access latency for both data load, store, and instruction fetch, bounding the worst-case latency when handling unpredictable events. %

Applications run on top of FreeRTOS~\cite{freertos}, an open-source, priority-based preemptive \gls{rtos}, in the manager core from where tasks are scheduled and run.

\subsection{Interrupts}\label{subsec:background-interrupts}
\subsubsection{Level- and edge-triggered interrupts}
Interrupt sources can signal interrupts either through a level change of the interrupt line, called a \textit{edge-triggered interrupt}, or by the logic level itself, a \textit{level-triggered interrupt}. While the latter requires the receiving side of the interrupt to clear the source often through accessing appropriate hardware registers, the former is unidirectional notification without confirmation. This results in faster processing but may lead to dropping interrupt requests accidentally.

\subsubsection{Vectored and direct interrupts}
Interrupts are asynchronous events that alter the \textit{normal program order execution} and require a switch to a different context to handle the event.
A processor supports \textit{vectored} interrupts when each interrupt traps to a specific \gls{isr} according to an \textit{interrupt vector table}, granting fast interrupt response at the cost of increased code size. Conversely, \textit{non-vectored} or \textit{direct} interrupts trap to a common \gls{isr}. The latter approach trades off code size for a slower interrupt response due to the overhead of resolving the cause of the interruption and jumping to the correct \gls{isr}.

\subsubsection{\textit{Interrupt latency}}\label{subsec:latency}
\begin{figure}[t]
    \centering
    \includegraphics[width=\linewidth]{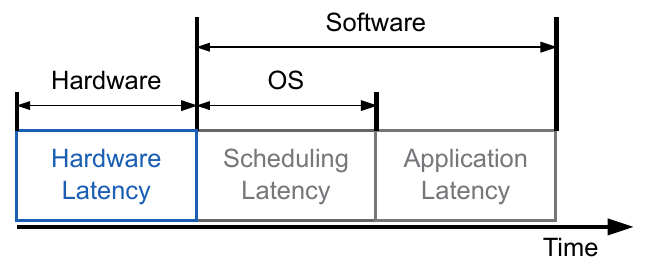}
    \caption{Interrupt latency breakdown into software and hardware dependent contributions.}
    \label{fig:int-latency-decomposed}
\end{figure}

Multiple sources determine the interrupt latency in a system: the underlying hardware, the scheduler/\gls{os}, and the application running on top as detailed in \cref{fig:int-latency-decomposed}.

In this paper, we focus on minimizing the latency imposed by the hardware; thus, whenever we refer to interrupt latency, we mean its hardware-contributed part.
Interrupt latency is defined as the time it takes from an interrupt edge arriving at the hardware, usually the interrupt controller, to the execution of the first instruction of the corresponding interrupt handler routine.
To make a fair comparison between software-based and more hardware-oriented interrupt solutions, we have to delineate what exactly constitutes the first instruction of the interrupt handler. We count as the first instruction the one after all necessary interrupt context has been saved on the stack to be able to call a \textit{function}. What exactly entails a function call is dictated by the used \gls{abi} or, more precisely, its calling convention. Note that when the interrupt handler and interrupt context saving code can be interleaved, some of the context saving code can be removed as it might be redundant.

\subsubsection{\textit{Context switching time}}
While interrupt latency is representative of the responsiveness of the hardware, scheduler, and application in arbitrating and servicing potentially asynchronous events with different priorities, context switching time refers to the responsiveness of the architecture in swapping from one execution context to another.
The execution context is dependent on the \gls{os} being used but usually consists of the \textit{state} of the architectural registers of the \gls{isa} and the chosen \gls{abi}.

\subsubsection{Nested and non-nested interrupts}\label{subsubsec:background-nested}
Preemption refers to an event, such as an interrupt request, temporarily interrupting a current task with the purpose of resuming its execution later.

The simplest case for preemption occurs with \emph{non-nested interrupt handlers}, where interrupts are globally disabled during the execution of an \gls{isr}. In the case of level/priority interrupt schemes, this means that (i) if the level/priority arbitration is \gls{sw}-driven, e.g., with priority simple/standard interrupt handlers, the highest priority interrupt identification code is not executed, and (ii) if the level/priority arbitration logic is designed within the interrupt controller, the highest level/priority interrupt is pending but disabled, hence not propagated to the core.
This scenario is not ideal for real-time and complex embedded systems, as interrupts are served sequentially. A high-priority interrupt has to wait for a lower-priority interrupt to finish.

A more complex case for preemption occurs with \emph{nested interrupt handlers} to handle the case of multiple interrupts at a time. In this scenario, interrupts are globally enabled within the scope of an executing \gls{isr}. The \gls{isr} need to care designed to ensure the they are reentrant.
With a level/priority interrupt scheme (either \gls{sw} or \gls{hw} driven), this introduces additional masking of incoming interrupts of equal or lower level/priority than the executing \gls{isr}, and sometimes larger than a configurable level/priority threshold. The nesting allows higher priority interrupts to preempt a current lower priority \gls{isr} executing.

\subsubsection{Redundant context restore with non-nested and masked back-to-back interrupts}\label{subsubsec:background-redundant-context-restore}
We indicate as \emph{back-to-back} interrupts served one after the other because preemption cannot occur. As detailed above, this is the case of interrupts globally masked in the non-nested scenario or masked according to level/priority schemes when nesting is allowed. In this situation, the active interrupt handler's context must be restored to be interrupted and saved again before handling the new interrupt event, causing a \emph{redundant context restore}. These redundant context restore sequences negatively impact interrupt latency on higher interrupt loads~\cite{lin_10}. \Cref{subsec:int_scenarios} explores the optimizations implemented in this work to address this scenario.


\subsection{Interrupt handling in RISC-V}\label{subsec:rv32-int}
\subsubsection{CLINT and PLIC}
We distinguish between \textit{core local interrupt controllers}, local to each \gls{hart}, and \textit{platform level interrupt controllers}, centralized interrupt controllers capable of addressing multiple \glspl{hart}.
In the RISC-V ecosystem, the privileged specification~\cite{RISCV_II} defines a simple interrupt scheme with a set of timer and inter-processor interrupts. We refer to it as the \gls{clint}. For 32-bit cores, it defines a fixed priority interrupt scheme with 16 predefined or reserved and 16 implementation-defined interrupts that can be optionally vectored.
The \gls{clint} itself only supports prioritization of interrupts based on privilege mode. This is insufficient for the embedded and real-time domain, and attempts to handle a more complex scheme would require software emulation, which incurs untenable interrupt latencies.
The interrupt handling is mostly software-based. To increase the number of custom interrupts, a \gls{plic}~\cite{riscv_plic} can be attached to the \gls{clint}. It can route a flexible (at design time) number of interrupts to one or more targets. Each interrupt can be assigned a priority, and each target can select a threshold below which interrupts are disabled. This scheme allows interrupts to be divided according to their priorities.

\subsubsection{CLIC}\label{subsubsec:background-clic}
The \gls{clic}~\cite{clic} addresses these limitations by allowing interrupts to be prioritized by so-called \textit{levels} and \textit{priorities}.
Interrupt selection is driven by the \gls{clic} in hardware (\cref{subsec:cv32rt_archi_clic}), which propagates the highest level, highest priority pending interrupt to the core's interface. Compared to standard RISC-V privileged specifications, pending and enabled interrupts are further selectively masked according to a \emph{threshold value} representing an interrupt level, configured through a \gls{csr}. Interrupts that are enabled, pending, and have a level below the threshold are masked, while others can be propagated. This feature is beneficial for \glspl{rtos} that allow certain types of critical interrupts to preempt execution still.

Consider two interrupts requests \texttt{irq1} and \texttt{irq2} with interrupt privilege modes, levels and priorities (\texttt{priv1}, \texttt{l1}, \texttt{p1}) and (\texttt{priv2}, \texttt{l2} and \texttt{p2}) respectively. 
Following the terminology introduced in \cref{subsec:background-interrupts} and by the \gls{clic}, we will refer to nested interrupts fired from different privilege modes as \textit{vertical} interrupts and to nested interrupts fired from the same privilege mode as \textit{horizontal} interrupts throughout this paper.
\cref{tab:clic-levels-prio} shows preemption conditions of two nested interrupts \texttt{irq2} and \texttt{irq1} according to the \gls{clic} specification. We consider an active interrupt handler servicing \texttt{irq2} while \texttt{irq1} is enabled, pending, and with a higher level than both the threshold and \texttt{irq2}.
Interrupt priority is considered a tie-breaker for the case of multiple horizontal interrupts with the same level and does not cause preemption, i.e., pending interrupts are serviced according to increasing priority.
Analogously, the case where multiple horizontal interrupts have equal levels and priorities results in the \gls{clic} selecting the highest numbered interrupt (identification number \texttt{id})~\cite{clic}, which is an arbitrary assignment at design time.

\begin{table}[t]
    \caption{Nested interrupt preemption scheme according to RISC-V \gls{clic}. We assume an interrupt request \texttt{irq2} is being serviced while another request \texttt{irq1} is enabled and pending. Preemption is regulated by privilege mode (\textit{vertical} interrupts) and interrupt level when the privilege mode is the same (\textit{horizontal} interrupts).}
    \label{tab:clic-levels-prio}
    \centering
    \renewcommand{\arraystretch}{1.5} 
    \resizebox{\linewidth}{!}{
    \begin{tabular}{c|c|cc}

           \thead{\textbf{Nested interrupt} \\ \textbf{behaviour}} & \textbf{Condition} & \thead{\textbf{CLINT} \\ \textbf{preemption?}} & \thead{\textbf{CLIC} \\ \textbf{preemption?}}  \\ \hline 

           \thead{Vertical} & \texttt{\textbf{priv1}} $>$ \texttt{\textbf{priv2}} & \cmark & \cmark\\ 
           
           \thead{Horizontal} & \thead{
           (\texttt{\textbf{priv1}} $==$ \texttt{\textbf{priv2}}) $\land$ \\  ((\texttt{\textbf{l1}} $>$ \texttt{\textbf{thresh}}) $\land$ (\texttt{\textbf{l1}} $>$ \texttt{\textbf{l2}}))
           } & \xmark & \cmark \\

    \end{tabular}
    }
\end{table}

Lastly, the \gls{clic} specification addresses the case of \textit{redundant context restore} (\cref{subsubsec:background-redundant-context-restore}) by introducing \texttt{xnxti}, a \gls{csr} short for \textit{Next Interrupt Handler Address and Interrupt-Enable CSRs} meant for use with non-vectored interrupts. Reading from this \gls{csr}, while the core is within an active handler, allows to fast-track interrupts that arrive late or to  avoid redundant context save/restore by running through pending interrupts back-to-back.

\section{Architecture}\label{sec:architecture}

Entering an interrupt context or performing a context switch requires the hardware to store enough information to resume operation correctly after returning from the aforementioned context. This needs to be done as fast as possible. From a high-level point of view, we can effectively improve interrupt latency and context switch times by:

\begin{itemize}
    \item Controlling the amount of state that needs to be preserved to enter and leave an interrupt context.
    \item Increasing the bandwidth and decreasing the latency to memory.
    \item Relying on \textit{latency-hiding} techniques that defer the effective saving of the state to a later point in time.
\end{itemize}

In the following, we detail the architectural features and \gls{hw}/\gls{sw} codesign of CV32RT (\cref{subsec:cv32rt_archi_overview} \cref{subsec:cv32rt_archi_clic}, and \cref{subsec:cv32rt_archi_fast}), and then describe how the proposed architecture tackles typical case scenarios (\cref{subsec:int_scenarios}). 

\subsection{CV32RT overview}\label{subsec:cv32rt_archi_overview}

We start with the CV32 core as the baseline, whose native \gls{clint} interrupt controller we replace with the \gls{clic} (hereafter, CV32RT$^{\texttt{CLIC}}$), and then introduce \texttt{fastirq} into the core microarchitecture, the extension proposed in this work (hereafter, CV32RT$^{\texttt{fastirq}}$). 
CV32RT$^{\texttt{fastirq}}$ optimizes interrupt latency in the non-nested interrupt case by combining bank switching and the nested interrupt case with an automatic context-saving mechanism in the background.
The background saving mechanism updates the stack pointer and stores the bank-switched contents in memory while the execution of the core proceeds in parallel.

We describe each of these changes in detail below.

\subsection{CV32RT$^{\texttt{CLIC}}$: CLIC fast interrupt controller}\label{subsec:cv32rt_archi_clic}
\begin{figure}
    \centering
    \includegraphics[width=\columnwidth]{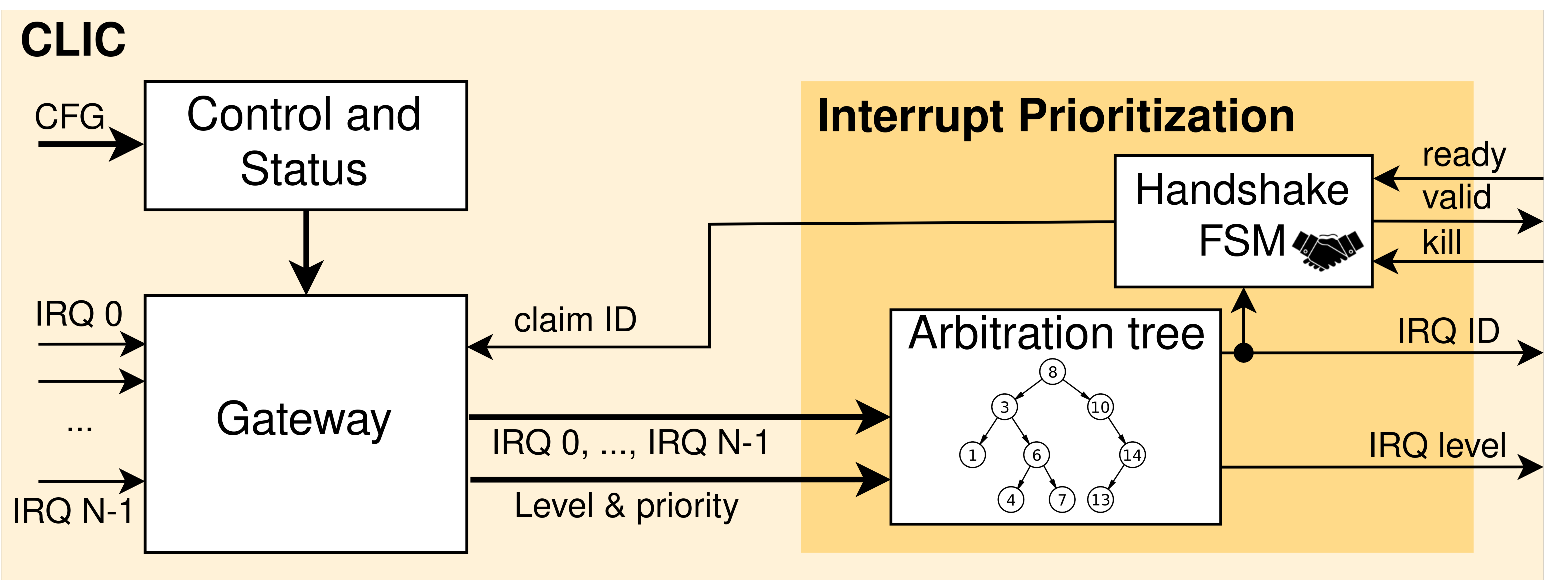}
    \caption{\gls{clic} architecture overview. Interrupts arrive at the gateway, where they are combined with
    programmable configuration information about each interrupt line consisting of level, priority, enable status, and sensitivity (level/edge). Enabled interrupts and their level and priority information are then further sent to prioritization logic which uses a binary arbitration tree to select the highest-level interrupt. The interrupt is then presented to the core with a handshake-based interface. The additional kill signal is there to allow for a handshake to restart so that a potentially more important interrupt can be presented to the core.}
    \label{fig:clic-arch}
\end{figure}

We implement the \gls{clic} interrupt controller in CV32RT$^{\texttt{CLIC}}$~\cite{clic} according to the draft specification which is to be included in the RISC-V Privileged Specification~\cite{RISCV_II}.
\cref{fig:clic-arch} provides an overview of the design. Incoming interrupts are filtered with a \texttt{Gateway} module that decides whether there is a pending and enabled request for each interrupt source \textit{i} (\texttt{IRQ i}). 

Assuming \textit{n} interrupt sources, selecting the interrupt with maximum level and priority is implemented with three binary trees. For each interrupt source, they track: (i) interrupt level and priority, (ii) interrupt \texttt{id}, and (iii) pending state from the gateway module. The \texttt{Interrupt Prioritzation} module traverses the tree from leaves to the root, where the sought-after maximum level and priority interrupt is found. 
Each tree has low overhead in terms of area and delay, $\mathcal{O}(n)$ and $\mathcal{O}(log(n))$, respectively.
The \gls{clic} design proposed in this work can scale up to $n=4096$ local interrupt sources. Optionally, additional pipeline stages can be inserted in the arbitration tree to relax timing.
Finally, our version of the \gls{clic} supports \gls{shv} and the \texttt{xnxti} \gls{csr} in the core.

Albeit improving the interrupt handling capabilities of CV32 by introducing priority and levels management in \gls{hw}, critical operations such as interrupt state and context save/restore are not natively covered by the \gls{clic} and need to be handled in \gls{sw}. CV32$^{\texttt{fastirq}}$ aims at filling this gap.

\subsection{CV32RT$^{\texttt{fastirq}}$: fast interrupt extension}\label{subsec:cv32rt_archi_fast}
\subsubsection{Automatic hardware context saving and bank switching}
\begin{figure*}
    \centering
    \includegraphics[width=\textwidth]{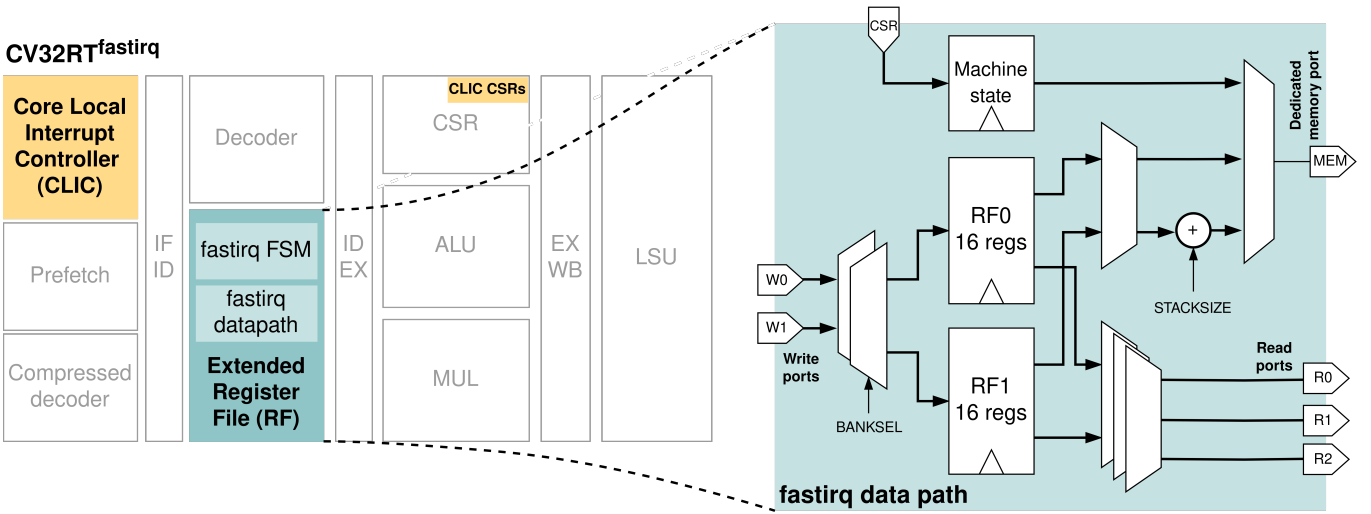}
    \caption{Overview of the fast interrupt architecture. The cv32's register file is extended with additional logic for the background saving mechanism. On an interrupt, the register file banks are switched. Parts of the old memory bank (the interrupt context) are copied to the core's stack location, while execution can go ahead by using the new bank. In this design, there is a dedicated memory port for the background saving mechanism, but optionally it can also be shared with the port from the load-store unit. The stack pointer is adjusted appropriately to maintain ABI invariants.}
    \label{fig:arch}
\end{figure*}

A block diagram of CV32RT$^{\texttt{fastirq}}$ is shown in \cref{fig:arch}. Conceptually, we can think of \texttt{fastirq} as a wrapper around the core's register file. We extend the register file by an additional read port for the background saving mechanism and registers for latching the additional processor state required for proper interrupt nesting.

A new interrupt at the \gls{clic} will first be checked whether the interrupt level exceeds the configured threshold. If this is the case, it will be dispatched to the core's main state machine while concurrently triggering the saving logic \gls{fsm} in the extended register file. The core's state machine will flush the pipeline and update the program counter according to the vector table entry. Meanwhile, in the extended register file, the saving logic triggers a bank switch, allowing the interrupt context to have a fresh set of registers while draining the other bank contents through a separate port to the main memory.

\subsubsection{Stack Pointer Handling}
During a bank switch, we need to update the stack pointer. For that, we have a dedicated adder between the two register files. The RISC-V embedded and integer ABI dictate that the stack pointer points below the last saved register on the stack. Furthermore, if we did not adjust the pointer, the program code running in the interrupt handler could clobber the values on the stack. We could do away with this mechanism for leaf-type interrupts, but this would require a new ABI considering a virtual stack pointer offset when generating code for interrupt handlers, and it would not solve the nested interrupt case.

\subsubsection{Forwarding Logic}\label{subsec:fwd}

While the interrupt handler can already be executed, we need to ensure that loads or stores access the stack values in memory since these are potentially still being written to memory by the background-saving mechanism. This is handled by informing the load-store unit of the core of the ongoing state being written to memory. A simple solution would be to block any memory accesses while the background saving mechanism is at work. This, though, partially negates the advantage of executing ahead since we likely want to issue a load soon in the interrupt handler.
In this implementation, we side-step this problem by checking whether the load-store unit tries to access the memory in the range of the stack pointer. The background saving happens incrementally. After the stack pointer is decremented, making space available, it will start storing the interrupt state word by word. In the load-store unit, the address offset of the last word pushed out by the background saving mechanism is compared against any incoming load and stores. Load and stores that try to access data that is not yet pushed to memory get stalled.

Alternatively, one could forward the value through the load-store unit instead, but this approach is more costly in terms of hardware complexity since this requires a dynamic address lookup into a queue-like buffer.

\subsubsection{Co-operation with Compiler}\label{subsubsec:compiler}

Accessing stack memory locations during the execution of an interrupt handler typically happens in two use cases:
\begin{itemize}
    \item A system call handler (issued through the \texttt{ecall} instruction in RISC-V) wants to access user-provided arguments. Most will be passed through general purpose registers, but some might be placed on the stack
    \item Short interrupt handlers that want to return before the full interrupt state has been saved.
\end{itemize}

In both cases, we do not need to engage the stalling logic outlined in \cref{subsec:fwd} by simply ordering the loads to first access the already stored interrupt state. This can be achieved since we know how the interrupt state gets pushed to memory. For example, if we push out the general purpose registers \texttt{x1, x2, \dots} in that very same order, then we need to ensure that in the interrupt handler, we use the same order to load words back.

Note that this is a performance issue and does not affect correctness. When writing assembly, the programmer has to be aware of that to achieve the best possible latency. In the case that the programmer uses compiler-specific attributes to write his interrupt handlers, such as \texttt{\_\_attribute\_\_((interrupt))} in GCC, the compiler needs to be made aware of that fact.

\subsubsection{Interrupt Handler Routines}
\begin{figure*}
    \includegraphics[width=0.9\textwidth]{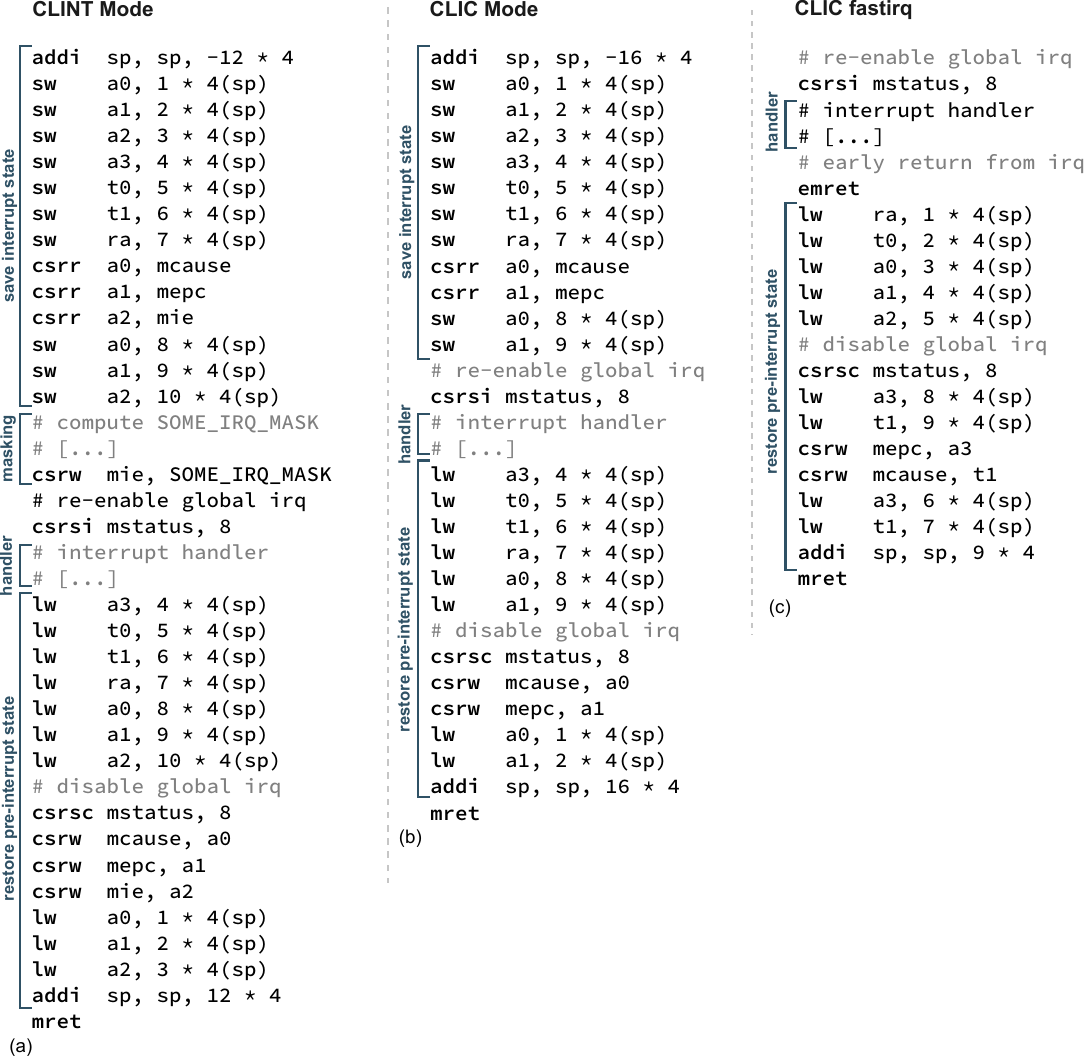}
    \caption{
Routines for saving state for vectored nesting interrupts using the \gls{clint} (a), \gls{clic} (b) and the proposed \texttt{fastirq} (c) extension. This assumes the regular RISC-V embedded \gls{abi}. Note that in \gls{clint}-mode interrupts with lower priority than the current interrupt running can fire when global interrupts are re-enabled. In order to prevent that, \texttt{mie} has to be manually adjusted such that the corresponding lower-priority interrupts are disabled.
    }
    \label{fig:nested-handler-listing}
\end{figure*}

In \cref{fig:nested-handler-listing}, we give an overview of how nested interrupt handling code works for the basic \gls{clint}-mode, the baseline \gls{clic}, and our \texttt{fastirq} extension. Due to the inflexible interrupt scheme of the \gls{clint}-mode, much more work needs to be done in managing interrupt mask (\texttt{SOME\_IRQ\_MASK}) and other machine state. The \gls{clic} addresses this with the level/priority scheme and \texttt{fastirq} improves upon that by moving the interrupt state saving logic in hardware and adding \texttt{emret} to handle redundant interrupt context sequences.

\subsection{Interrupt scenario analysis}\label{subsec:int_scenarios}
\subsubsection{Non-nested interrupts}
As discussed in \cref{subsec:background-interrupts}, redundant context restore with non-nested or nested horizontal interrupts can introduce unwanted additional interrupt latency.
With the proposed approach, the interrupt state can be quickly restored by simply switching register banks. To differentiate between a regular return from an interrupt handler using \texttt{mret}, which assumes the interrupt state has been restored by software, we add a new instruction \texttt{emret} instead. This instruction performs the same function as \texttt{mret} in addition to switching register banks.

In the RISC-V ecosystem, this situation can be optimized by directly checking for other interrupts pending on the same level before restoring the execution's interrupt context. 
The \gls{clic} offers a hardware-assisted solution to address such a scenario with the \texttt{xnxti} \glspl{csr} (\cref{subsubsec:background-clic}). Nuclei's ECLIC~\cite{nuclei_isa} extends \texttt{xnxti} by embedding the jump to the queuing interrupt handler in the \texttt{xnxti} hardware (\texttt{jalxnxti}).
In our case, this is implemented through the \texttt{emret} instruction. Besides checking whether a bank switch is appropriate, it also looks at interrupts pending with the same or lower level. If such an interrupt exists, \texttt{emret} redirects the control flow to the pending interrupts handler. Some hardware refers to this concept of removing \textit{redundant context restores} as tail-chaining~\cite{yiu_cortex_reference_13}. This situation is visualized in \cref{fig:nested-interrupt-example}(b).

\subsubsection{Nested interrupts}
\begin{figure}[t]
    \centering
    \includegraphics[width=\columnwidth]{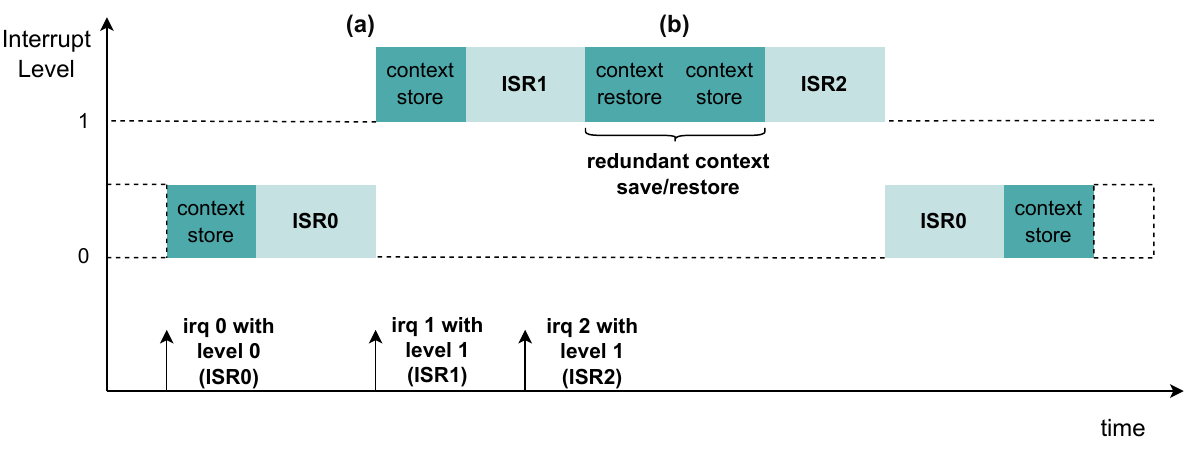}
    \caption{\textbf{(a)} Transition between different interrupt levels. \textbf{(b)} Redundant context restore of two non-preemptive interrupts (e.g., two interrupts with same level but different priorities) and tail-chaining to optimize it.}
    \label{fig:nested-interrupt-example}
\end{figure}

Following the notation adopted in \cref{subsubsec:background-nested}, the case of nested interrupts entails preemption of a low-level interrupt by a high-level interrupt, as shown in \cref{fig:nested-interrupt-example}(a). We address this scenario by minimizing \textit{interrupt latency} through hiding latency.

Whenever an interrupt handler is entered, global interrupts are disabled. This increases the worst-case interrupt latency by the time they remain disabled. For nesting interrupts, the first instruction will be re-enabling global interrupts. At this point, a high-level interrupt could preempt the current running interrupt handler. If this happens while the background saving mechanism is still at work, the execution of the handler has to wait.

This combination of latency hiding and background saving allows the core to quickly enter a first-level interrupt handler and potential nesting interrupt handlers, albeit with a larger delay.

Restoring the pre-interrupt context is entirely handled in software for the nested interrupt case. While this return path could also be handled in hardware by adding an additional write port to the core's register file, exiting an interrupt handler is less time-critical since we mostly care about how quickly an external event is addressed.

\subsection{Context Switching acceleration}
We can divide context switches into operating system-specific and hardware-specific parts. The operating system part entails all contributions to the context switch time that is specific to the operating system itself, such as computing the next task to be scheduled and bookkeeping operations. The remainder is the hardware-dependent saving and restoring of the state belonging to the new context.

The idea is to use the background saving mechanism to accelerate the state saving and restoring part of context switches by interleaving the loading of a new state with the automatic saving hardware pushing out the previous register state to memory in the background.

For that, the initial part of the context switch routine changes. At the beginning of the routine, we want to save the current running tasks' state to memory. Instead of manually saving the general purporse registers, we set up the stack pointer appropriately before triggering a software interrupt (write to \gls{clic}'s memory map). This will engage the hardware mechanism to swap the registers and save them in the background while we can already proceed with the rest of the context switch routine.

\section{Evaluation}\label{sec:evaluation}

This section gives a functional and quantitative evaluation of the various flavors of the CV32RT. We give an analysis of the functional improvements and argue that these allow more flexible and efficient handling of regular and nested interrupts. Then we show how the various CV32RT versions perform in terms of interrupt latency and context switch times and finally quantify the overhead these additions incur in terms of area and timing.

\subsection{Feature Comparison}
As discussed in \cref{sec:background}, the \gls{clint} is the first standardized RISC-V interrupt controller. It supports pre-emption based on privilege modes (machine, hypervisor, supervisor, user), but the interrupt lines have a hardwired prioritization scheme.
For the embedded use cases, the \gls{clint} lacks fine-grained control over interrupt prioritization. The baseline \gls{clic} addresses these weaknesses.

Interrupts can be dynamically assigned a priority and a level. The priorities allow concurrent pending interrupts to be taken in the order preferred by the programmer, while the level information enables pre-emption of same-privilege level interrupts (also called horizontal interrupts). Interrupts that are assigned a higher level can pre-empt lower-level interrupts. In addition to that, a level threshold register per privilege level (\texttt{xintthresh}) controls the set of allowed horizontal interrupts by limiting it to those whose's level exceeds the given value in the register. This can be useful in critical sections where only a subset of interrupts need to be disabled.

To improve interrupt latencies, interrupt vectoring is supported. Vectoring can be selectively enabled or disabled per interrupt line. This allows for better control of the vector table size.
The baseline \gls{clic} does not differ from the regular \gls{clint} in terms of storing and restoring the interrupt context which is fully handled in software.

The optional \texttt{xnxti} extension allows multiple horizontal interrupts to be serviced in sequence without redundant context-restoring operations in between. Reading the \texttt{xnxti} \gls{csr} yields a pointer to the vector table entry for the next pending and qualifying interrupt, if available, which then allows a direct jump there. This approach loses the latency advantage of hardware vectoring by basically running a software emulation thereof. In addition, the first interrupt has to still pay the full latency cost since \texttt{xnxti} does not touch the interrupt context storing part itself.

Our proposed \texttt{fastirq} extension addresses these weaknesses by extending the \gls{clic} base capabilities with a mechanism to lower interrupt latency while still keeping hardware vectored interrupts and allowing the skipping of redundant context restore operations.

The discussed differences are summarized in \cref{tab:functional}.

\begin{table}[t]
    \caption{Feature comparison of various flavors of RISC-V core-specific interrupt controllers.}
    \label{tab:functional}
    \centering

    \renewcommand{\arraystretch}{1.3} 
    \resizebox{\linewidth}{!}{
    \begin{tabular}{l|c|c|c}
           \textbf{Configuration}                    & \thead{\textbf{Prioritization by} \\ \textbf{\textit{priv.}| \textit{level} | \textit{priority}}} & \thead{\textbf{Vectored} \\ \textbf{Interrupts?}} & \thead{\textbf{Redundant} \\ \textbf{Context Restore}} \\
           \hline
           CV32 (\gls{clint})                      & \cmark | \xmark | \xmark           & Both   & \xmark \\
           CV32RT$^{\texttt{CLIC}}$                       & \cmark | \cmark | \cmark           & Both   & \xmark \\
           CV32RT$^{\texttt{CLIC}}$ (\texttt{xnxti})   & \cmark | \cmark | \cmark           & \xmark & \cmark \\
           CV32RT$^{\texttt{CLIC}}$ (\texttt{jalxnxti}) & \cmark | \cmark | \cmark           & \xmark & \cmark \\
           CV32RT$^{\texttt{fastirq}}$ & \cmark | \cmark | \cmark           & \cmark & \cmark
    \end{tabular}
    }

\end{table}

\subsection{Functional performance}

\subsubsection{Measurement Setup}
We take measurements by running RTL simulations of the different versions of CV32RT as part of ControlPULP. For the memory subsystem, this means we have single cycle (zero wait state) access to \gls{sram} (\cref{subsec:platform}). The memory bank we are using is not contended by other bus masters which is achieved by placing data and instructions into specific private banks. On the system level, there are no additional latencies introduced on interrupt lines between interrupt sources and the \gls{clic}.

\subsubsection{Interrupt Latency}\label{subsubsec:arch-int-latency}

We measure the hardware contributed interrupt latency (as described in \cref{subsec:latency}) of our \texttt{fastirq} extension and compare it to the CV32 and CV32RT variations (standard CLIC, \texttt{xnxti}, \texttt{jalxnxti}). The interrupt latency is measured as the number of cycles it takes for an interrupt to arrive at the interrupt controller input to the first instruction of an interrupt handler that allows the calling of a C-function. This implies that all caller-save registers need to be saved, which is the general case.

Interrupt handler routines that save the interrupt context in software can only save the minimum state if the compiler is able to fully inline the handler's function body. Depending on the function and the ABI, this might be fewer than all caller-save registers. To also consider these cases, we measure the interrupt latency in the optimal case, i.e., only one caller-save register needs saving for software-based interrupt handlers.

Interrupts are injected into the design at the interrupt controller inputs. Each hardware configuration has a handwritten optimized interrupt handler that stores all required general-purpose and machine-specific registers for nesting interrupts. We evaluate both the EABI and regular integer ABI of RISC-V. The saved general-purpose registers are all caller-save registers in the respective ABI. This allows for directly calling into C-code functions from interrupt handlers. For our \texttt{fastirq} extension, this amount of state is always saved since the hardware does not know a priori which caller-save registers need to be saved. Interrupt handler routines that use software-based mechanisms to save and restore interrupt state can, if the compiler permits, fully inline the handler code and save some caller-save registers. 

The results are summarized in \cref{fig:performance}. Compared to the baseline \gls{clint} and \gls{clic} which both have about 33 cycles interrupt latency, the \texttt{fastirq} extension is able to reduce this down to 6 cycles. \texttt{xnxti} and \texttt{jalxnxti} perform even worse (requiring 42 and 35 cycles, respectively) since they both insert additional instructions in the code path between the handler and interrupt event. As a comparison, the Arm Cortex-M4 has an interrupt latency of 12 cycles given a single-cycle memory~\cite{cortex_m4_manual}.

\subsubsection{Redundant Context Restoring}
As described in \cref{subsec:background-interrupts}, whenever there are \emph{back-to-back} interrupts, the interrupt context restore and store sequence between them can be considered redundant.

\begin{figure}[t]
    \centering
    \includegraphics[width=\columnwidth]{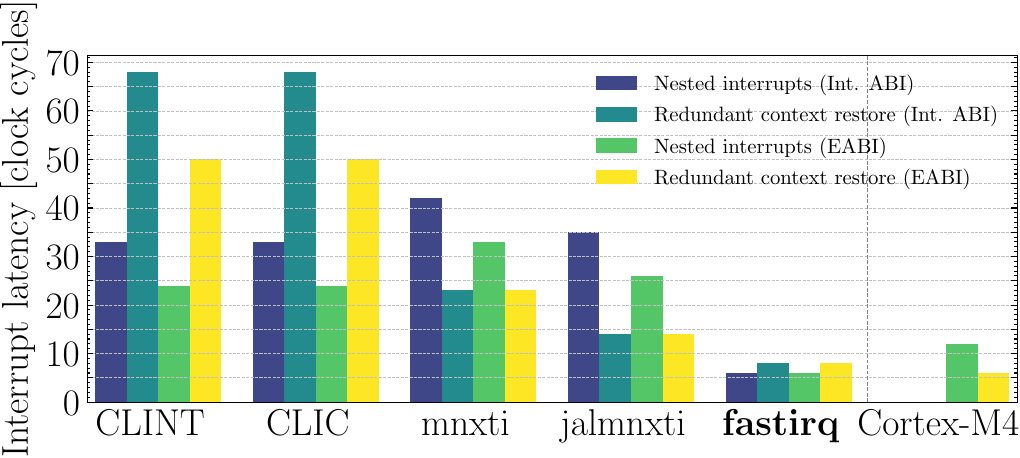}
    \caption{Interrupt latency of various flavors of RISC-V core-specific interrupt controllers for interrupt handlers that support nesting and the calling of C-functions within it (i.e., saving/restoring state following the C-ABI). As a comparison, we mention the generic Cortex-M4 core, assuming it has single-cycle access to memory. Note that Arm Cortex-M only has 16 core registers.}
    \label{fig:performance}
\end{figure}

In \cref{fig:performance} we show the cost in clock cycles of such sequences. For the baseline \gls{clic} it takes 68 or 50 cycles when using the integer or embedded ABI respectively. In this case, the redundant context restoring sequences contains the full interrupt exit code sequence. The latter consists of restoring the interrupt context and the regular interrupt latency. \texttt{xnxti} and \texttt{jalxnxti} improve upon this situation by checking for pending interrupts and directly jumping to the respective handlers. Note that this is independent of the ABI since no interrupt state is affected. While \texttt{xnxti} only returns a pointer to the address of the next handler and thus needs a small code sequence (load, jump, and retry loop), \texttt{jalxnxti} fuses these operations into one instruction resulting in saving 9 cycles. The \texttt{emret} mechanism of \texttt{fastirq} works similarly and costs 8 clock cycles but is not restricted to non-vectored interrupts. The Arm Cortex-M4 is able to do the same task in 6 cycles assuming it has access to single-cycle memory~\cite{cortex_m4_manual}.

\subsubsection{Context Switch Time}\label{subsubsec:arch-int-context-switch}
\begin{figure}[t]
    \centering
    \includegraphics[width=\columnwidth]{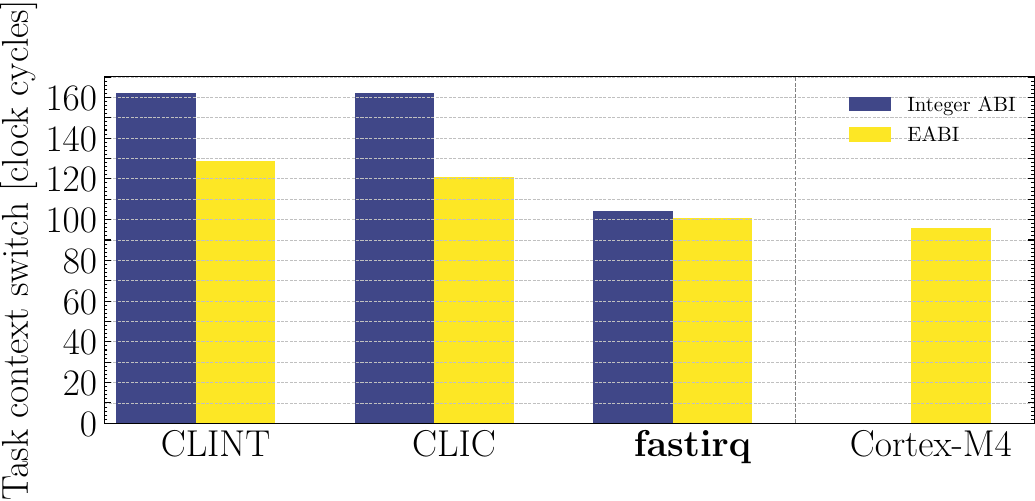}
    \caption{Average Context Switch Time in FreeRTOS for two tasks of various flavors of CV32RT. The RISC-V E-Extension reduces the available general purpose registers from 32 to 16, consequently lowering the context switch state that needs to be saved and restored.}
    \label{fig:context_switch}
\end{figure}

In \cref{fig:context_switch} we show the context switch time in number of clock cycles between two FreeRTOS dummy tasks when comparing the baseline CV32RT against CV32RT$^\texttt{fastirq}$. All compile time options such as tracing, stack overflow signaling, and the more generic task selection mechanism were turned off to minimize the context switch code. We left out configurations that do not have any impact on context switching. \texttt{fastirq} allows us to skip ahead the saving of the general purpose registers by triggering a software interrupt as part of the save sequence. The software interrupt will trigger the \texttt{fastirq} mechanism, which starts saving the general purpose registers to memory. By making sure not to use registers that are still being saved by the background saving mechanism (as described in \cref{subsubsec:compiler}) we can save up to 31 cycles ($-19\%$) for a context switch when using the I-extension, and 16 cycles ($-12\%$) for the E-extension, respectively. The FreeRTOS website~\cite{freertos} claims context switches as low as 96 cycles for a Cortex-M4 implementation.~\footnote{This can be seen as a lower bound as on some \glspl{soc} such as the STM32L476RG we observed higher latencies due to memory access stalls and other implementation choices in the memory subsystem.}

\subsection{Implementation}
We synthesize CV32RT as part of the ControlPULP platform using Synopsys Design Compiler 2022.03, targeting GlobalFoundries 12LP FinFet technology at \SI{500}{\mega\hertz}, TT corner, and \SI{25}{\celsius}.
One gate equivalent (GE) for this technology equals 0.121 $\mu$m$^2$.

\begin{figure}[t]
    \centering
    \includegraphics[width=\columnwidth]{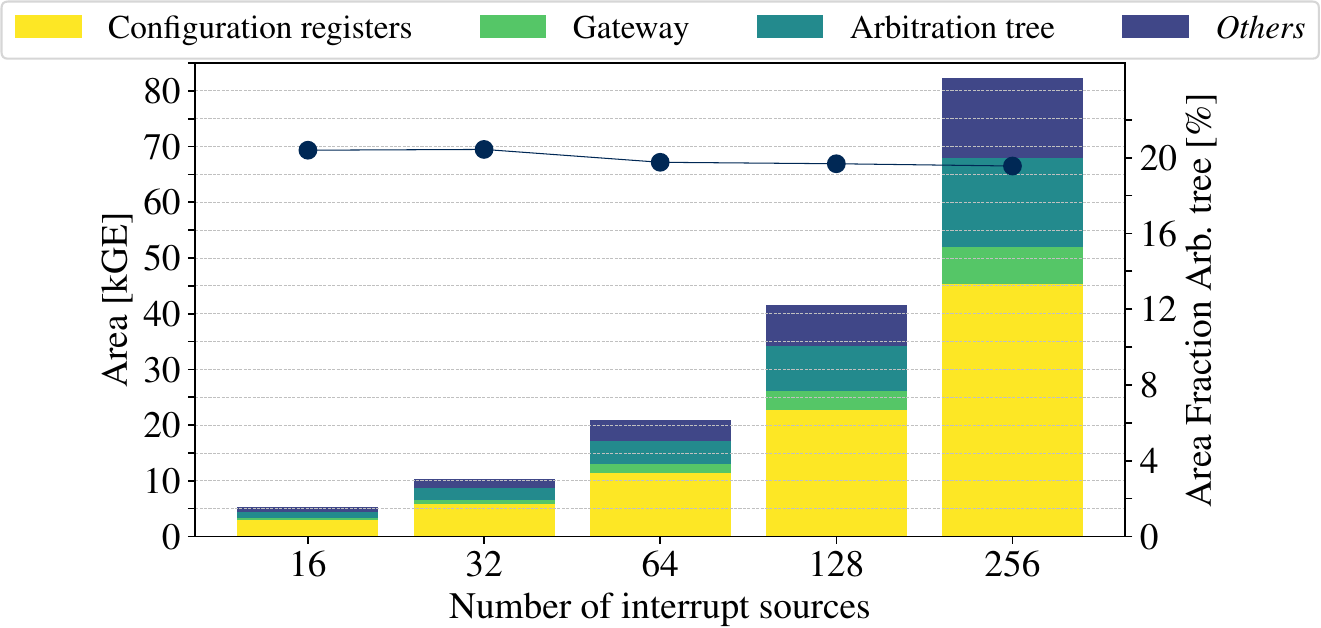}
    \caption{RISC-V \gls{clic} area breakdown at varying numbers of interrupt sources.}
    \label{fig:clic-breakdown}
\end{figure}

\begin{figure}[t]
    \centering
    \includegraphics[width=\columnwidth]{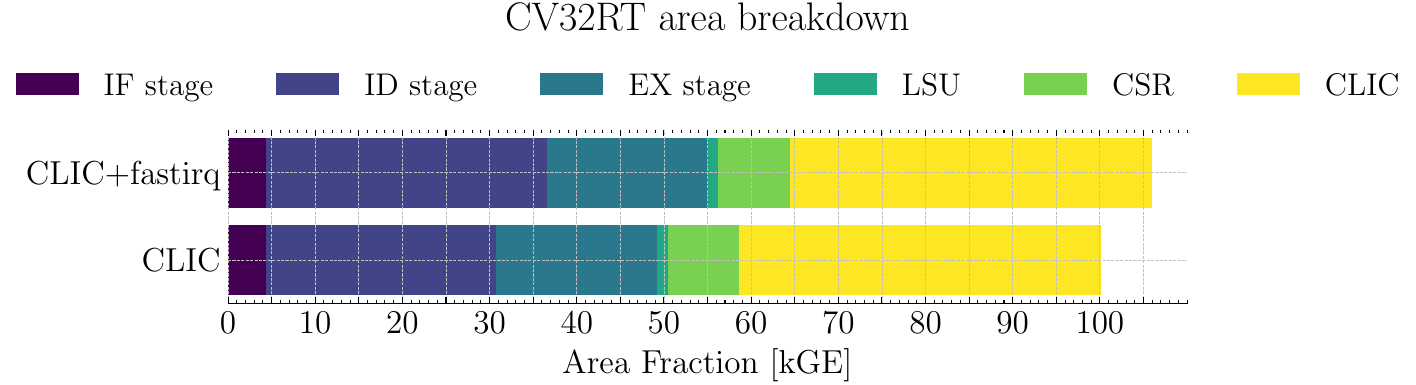}
    \caption{Area overhead of CV32RT in the two main configurations.
    The overhead of \texttt{fastirq} in CV32RT$^{\texttt{fastirq}}$ core results in a minimal 10\% area increase concentrated around the ID stage.
    The design has been synthesized in GF12LP technology (500 MHz, \textit{TT} corner, \SI{25}{\celsius}, 0.8V, super low V$_T$ standard cells).}
    \label{fig:cv32rt-breakdown}
\end{figure}

\begin{figure}[t]
    \centering
    \includegraphics[width=\columnwidth]{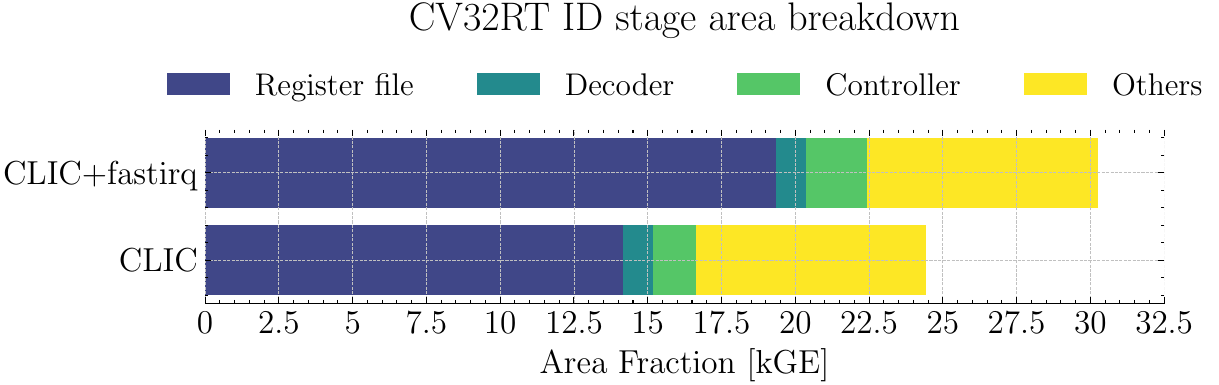}
    \caption{Area breakdown of CV32E40P$^{RT}$ \textit{ID stage} with the proposed hardware extensions.
    }
    \label{fig:cv32rt-id-breakdown}
\end{figure}

\Cref{fig:clic-breakdown} reports the area breakdown of the \gls{clic} implemented in the proposed with different interrupt sources.
As the figure shows, for each scenario, more than half of the resources implement the configuration registers required to control the \gls{clic}, which size linearly increases with the number of input interrupts. 
In particular, each interrupt is associated with one 32-bit register incurring an area overhead of about \SI{176}{GE}. The remaining area implements the gateway and binary tree arbitration logic at the core of the \gls{clic} working principle, as discussed in \cref{subsec:cv32rt_archi_clic}, and additional housekeeping control logic that scales linearly with the number of interrupt sources.
\Cref{fig:clic-breakdown} also shows that the fraction of the design occupied by the arbitration tree is kept constant when increasing the number of sources.
While it is apparent that this design incurs a larger area overhead compared to traditional RISC-V \gls{clint}~\cite{RISCV_II}, the gain in flexibility enables a broader application scope with time-critical systems.

\Cref{fig:cv32rt-breakdown} reports the area breakdown comparison of CV32RT with baseline \gls{clic} (including \texttt{mnxti} and \texttt{jalmnxti} \glspl{csr}) and \texttt{fastirq} extensions with 256 input interrupts.
From the table, CV32RT$^{\texttt{fastirq}}$ incurs a 10\% overall area increase compared to baseline \gls{clic} only.
While other \gls{hw} blocks of the core remain primarily unaffected, the \gls{id} stage incurs an area overhead of 21\% compared to CV32RT$^{\texttt{CLIC}}$. 
The \gls{id} stage is the \gls{hw} block where the additional registers and the automatic stacking/unstacking logic are localized. 
A breakdown of the \gls{id} stage is shown in \cref{fig:cv32rt-id-breakdown}.
We notice that the additional storage space for automatic context save and restore in \gls{hw} increases the area of the register file by about 36\% in the proposed implementation. Similarly, the logic for managing the shadow registers accounts for an overhead of 40\% on the baseline \textit{ID stage} controller. The \gls{hw} overhead coming from the additional \texttt{emret} instruction discussed in \cref{sec:evaluation} is negligible.
Nevertheless, the increased size of the \gls{id} stage trades off the benefits of (i) a simplified programming model that moves several \gls{sw} operations in \gls{hw}, and (ii) significantly lowered interrupt latency than standard RISC-V while not impacting the critical path of the base core design. 
Lastly, time-critical systems shift design priorities from area efficiency to safety, security, and reliability.

\section{Related Work}\label{sec:related_works}

In this section, we describe the leading solutions to optimize handling asynchronous events in \gls{sota} embedded and real-time \glspl{mcu}.
We first differentiate between existing platform-level and core-local interrupt controllers as introduced in~\cref{subsec:rv32-int}. Then, we focus on the latter and discuss solutions across various platforms in industry and academia, addressing (i) interrupt context save/restore techniques, (ii) context switch techniques, and (iii) dedicated strategies to optimize redundant context restore with back-to-back interrupts.
\Cref{tab:related_work} summarizes the overview. 
For the interrupt latency, we assume the definition presented in \cref{subsec:background-interrupts} and explicitly provide references to different variants adopted by \gls{sota} when needed in the table.

\begin{table*}[t]
    \caption{Comparison of the main techniques for optimizing interrupt context and task context save/restore with nested and non-nested interrupts employed by industry and academia in the embedded and real-time application domains.}
    \begin{center}
    \setlength{\tabcolsep}{10pt} 
    \renewcommand{\arraystretch}{1.4} 
    \resizebox{\linewidth}{!}{
    \begin{threeparttable}
    \begin{tabular}{cccccc|ccccc}
    \rotatebox{0}{\textbf{Processor}} & 
    \rotatebox{0}{\textbf{\thead{Interrupt \\ controller}}} & 
    \rotatebox{0}{\textbf{\gls{isa}}} & 
    \rotatebox{0}{\textbf{\thead{Interrupt context \\ save/restore acceleration}}} & 
    \rotatebox{0}{\textbf{\thead{Task context \\ save/restore acceleration}}} & 
    \rotatebox{0}{\textbf{\thead{Back-to-back interrupts \\ acceleration}}} & 
    \rotatebox{0}{\textbf{Open?}} & 
    \rotatebox{0}{\textbf{\thead{Max. \\ \#levels}}} &
    \rotatebox{0}{\textbf{\thead{Interrupt \\ latency}}} & 
    \rotatebox{0}{\textbf{\thead{Task context \\ switch time}}} \\


    \hline 
    \multicolumn{10}{l}{\textbf{Industry}} \\ 
    \hline 
    
    Arm Cortex M4~\cite{arm_interrupt_beginner_guide, freertos} & NVIC & Arm RISC & Automatic in \gls{hw} & n.a. & NVIC tail-chaining & \xmark & 256 & 12 & 96 \\ 
    Arm Cortex R5 & VIC/GIC & Arm RISC & n.a. & Register banking & n.a. & \xmark & n.a. & 20 & n.a. \\
    Infineon AURIX & ICU & TriCore RISC & Automatic in \gls{hw} & \Gls{sw} based - \gls{csa} & n.a. & \xmark & 256 & 10-16 & 156-162 \\
    Renesas M32C/80 & IC & M32C/80 & Dual register bank & Dual register bank & n.a. & \xmark & 8 & 5~\tnote{a} & n.a. \\
    Nuclei System Technology & ECLIC & RISC-V & n.a. & n.a. & \texttt{jalmnxti} & \xmark & 256 & 4-6~\tnote{b} & n.a. \\
    Alibaba’s T-Head Xuantie E906 & CLIC & RISC-V & n.a. & n.a. & \texttt{mnxti} & \cmark & n.a. & n.a. & n.a. \\
    SiFive E21 & CLIC & RISC-V & n.a. & n.a. & \texttt{mnxti} & \xmark & 16 & 20~\tnote{c} & n.a. \\
    ESP32 C3 & INTC & RISC-V & n.a. & n.a. & n.a. & \xmark & 15 & n.a. & n.a. \\
    - & AIA local & RISC-V & - & - & \texttt{xtopi} & \xmark & - & - & - \\ 


    \hline 
    \multicolumn{10}{l}{\textbf{Academia}} \\ 
    \hline 
    
    Gaitan et al.~\cite{TASK_SWITCHING_TVLSI} & MPRA/HSE & RISC & Automatic in \gls{hw} & Automatic in \gls{hw} & n.a. & \xmark & n.a. & 1.5~\tnote{d} & 3~\tnote{d} \\
    Mao et al.~\cite{mao_2021} & CLIC & RISC-V & Automatic in \gls{hw} & n.a. & n.a.~\tnote{e} & \xmark & n.a. & 13 & n.a. \\
    Balas et al.~\cite{balas_2021} & CLINT & RISC-V & - & - & - & \cmark & - & 24 & 129 \\
    \textbf{This work} & \textbf{CLIC} & \textbf{RISC-V} & \textbf{Hybrid}~\tnote{\textbf{f}} & \textbf{Hybrid} & \texttt{\textbf{emret}} & \cmark & \textbf{256} & \textbf{6} & \textbf{114} \\
    \hline   
    \end{tabular}
    \begin{tablenotes}
    \item[a] Single high-speed interrupt. Refers to the number of clock cycles from the firing of an interrupt to the execution of the interrupt body. Does not count the saving of the general purpose registers.~\cite{renesas_m32c80}.
    \item[b] Refers to the number of cycles from arrival the arrival of the interrupt to the first instruction fetch of the \gls{isr}~\cite{sifive_e21}. Does not count the saving of the interrupt context, including the general purpose registers.
    \item[c] Refers to the number of the clock cycles it takes from the firing of an external interrupt to the time the first instruction in the corresponding interrupt service routine of a C function is executed~\cite{nuclei_isa}.
    \item[d] Full \gls{hw} solution. \Gls{sw} scheduling latency becomes negligible as from~\cref{fig:int-latency-decomposed}, which justifies the performances in the order of the clock cycle.
    \item[e] Implementation details are omitted.
    \item[f] Combination of background saving/restoring and register banking.
    \end{tablenotes}
    \end{threeparttable}
    }
    
    \label{tab:related_work}
    \end{center}
\end{table*}

\subsection{Platform-level interrupt controllers}
Controllers that belong to this category, such as the RISC-V \gls{plic} and \gls{aplic} for wire-based interrupt communication and RISC-V \gls{imsic} for message-signaled interrupt communication, are not designed to distribute time-critical interrupts to the running \glspl{hart}. 
Conversely, Arm's \gls{gic} redistributes incoming asynchronous events as either non-critical (\texttt{IRQ}) or critical interrupts (fast \texttt{IRQ}, or \texttt{FIQ}).
The latter reduces the execution time of the interrupt handler through a dedicated register bank, with up to eight registers employed to minimize context switching.

\subsection{Core-local interrupt controllers}
Core-local interrupt controllers are often specialized in providing fast interrupt-handling capabilities in real-time embedded application domains.

\subsubsection{Interrupt context save/restore acceleration}
Several designs automatically save and restore the interrupt context directly in \gls{hw}. This allows the core to simultaneously start fetching the jump address from the interrupt vector table during context save and embed context restore within the return instruction.
A reference example in the field is the Arm Cortex-M series integrating the \gls{nvic}. It implements a state machine~\cite{stm32l5} that performs caller-save register stacking in the background. Likewise, upon returning from the \gls{isr}, the \gls{hw} encodes in the link register a value (\texttt{EXC\_RETURN}), which notifies the core to start unwinding the stack to return to normal program execution.
Analogously, in Infineon AURIX \gls{mcu}-class TriCore family's \gls{icu}~\cite{tricore_hotchips16, aurix_tc27x, aurix_tricore_v1.6} the context of the calling routine is saved in memory autonomously while restoring the context is embedded in the \texttt{RET} instruction and happens in parallel with the return jump~\cite{tricore_patent}.
From academia, Mao et al.~\cite{mao_2021} are the first to propose extensions for the RISC-V \gls{clic}. Though implementation details are omitted, the interrupt handling is enhanced with automatic stacking in hardware that benefits from the core's Harvard architecture and simultaneous data and instruction memory access.

Albeit automatic interrupt context save/restore reduces \gls{sw} housekeeping overhead before and after handling the interrupt routine, it only partially addresses the acceleration of the complete task context switch.

\subsubsection{Task context save/restore acceleration}
Register banking is a technique adopted by several interrupt controller architectures to swap a task's context without pushing/popping register values to the stack at the cost of an additional area overhead in the design.
As discussed above, in Arm designs, this is the case of \gls{hw} register banking in platform-level interrupt controllers (\gls{gic}). 
A similar approach is implemented in the Renesas M32C/80 series~\cite{renesas_m32c80}. A dual register bank allows quickly swapping the context without saving/restoring to/from the stack, as the second register bank is reserved for high-speed interrupts.
The Aurix family instead implements a \gls{sw} managed solution, featuring a specific organization of the context layout in the system memory based on \glspl{csa} chained in a linked list fashion~\cite{tricore_patent}.
These approaches trade-off \gls{hw} and \gls{sw}-induced latencies when handling asynchronous events, as depicted in~\cref{fig:int-latency-decomposed}. Gaitan et al.~\cite{TASK_SWITCHING_TVLSI} propose instead a full \gls{hw} solution based on a Hardware Scheduling Engine (HSE) that directly attaches interrupts to running tasks without the need for a specialized interrupt controller. While this approach allows lowering interrupt latency and task context switches dramatically, as reported in \cref{tab:related_work}, it is expensive in terms of area as it requires replicating hardware resources per task.

\subsubsection{Optimizations for non-nested interrupt handling}
Redundant context restore with non-nested interrupts is addressed by chaining two back-to-back interrupts and bypassing the superfluous restore/save operation. 
RISC-V addresses this scenario by adding the \texttt{xnxti} \gls{csr} with \gls{sw}-managed interrupt service loops as part of the \gls{clic} specifications and finds early commercial examples with~\cite{sifive_e21}. 
Nuclei System Technology \textit{enhanced \gls{clic}} (ECLIC)~\cite{eclic} extends traditional \texttt{xnxti} with a novel \gls{csr} for machine privilege mode, \texttt{jalmnxti}~\cite{nuclei_isa}, already discussed in this work in \cref{sec:evaluation}.
Among MSI solutions, RISC-V AIA without APLIC has a similar approach to \texttt{mnxti} with the \texttt{xtopi} \gls{csr} that reports the highest-priority, pending, and enabled interrupt for a specific privilege mode, allowing both late arrival and redundant context restore mechanisms.

While the presented solutions are effective in optimizing context, save/restore with \gls{hw} and \gls{sw} cooperation, they lack (i) a cohesive approach to address both interrupt context and task context switch acceleration and (ii) none of the existing RISC-V-based approaches can close the gap with well-established industry vendors.

This work proposes a combination of the background-saving with a register banking approach. Interrupt context save/restore especially takes advantage of the former by deferring those operations to the \gls{hw}. This allows execution of the interrupt handler before all interrupt state has been pushed to memory, lowering interrupt latency.
A task's context switch benefits from quickly transferring the suspended context to the dedicated register bank while already restoring the next task to be executed.

To address redundant context restore with back-to-back interrupts, we propose a modified version of the return instruction in machine mode (i.e., \texttt{emret}), which is able to skip redundant context saving and restoring sequences by directly jumping to the next available interrupt handler. This solution works for both vectored and non-vectored interrupts.

\section{Conclusion}\label{sec:conclusion}
In this work, we present \texttt{fastirq}, a fast interrupt extension for RISC-V embedded systems. We implement the extension on CV32RT, a 32-bit, in-order, single-issue core designed with the RISC-V \gls{clic} fast interrupt controller.
With our design, we can achieve interrupt latencies of 6 clock cycles and efficient back-to-back interrupt handling in 12 cycles which is as low as the fastest available approaches currently implemented in the RISC-V landscape, fully open-source, and competitive against closed-source and proprietary commercial solutions. Furthermore, we improve task context switch times in FreeRTOS to 104 clock cycles using \texttt{fastirq}, which is $20\%$ faster than a software-only approach.


\bibliographystyle{bibliography/IEEEtranTIE}
\bibliography{bibliography/main}\ 

\newcommand{\lucaphd}{He is currently pursuing a Ph.D. degree in the Digital Circuits and Systems group of Prof.\ Benini.}
\newcommand{\ethgrad}[2]{received his BSc and MSc degrees in electrical engineering and information technology from ETH Zurich in #1 and #2, respectively.}
\newcommand{\researchinterests}[1]{His research interests include #1.}

\begin{IEEEbiography}[%
    {\includegraphics[width=1in,height=1.25in,clip,keepaspectratio]%
        {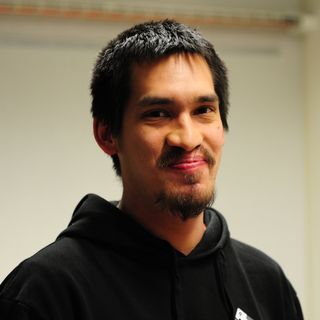}}%
    ]{Robert Balas}
    \ethgrad{2015}{2017}
    \lucaphd{}
    \researchinterests{real-time computing, compilers, and operating-systems}
\end{IEEEbiography}

\begin{IEEEbiography}[%
    {\includegraphics[width=1in,height=1.25in,clip,keepaspectratio]%
        {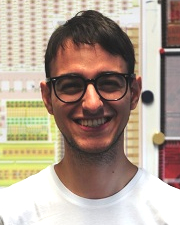}}%
    ]{Alessandro Ottaviano}
    received the B.Sc. in Physical Engineering from Politecnico di Torino, Italy, and the M.Sc. in Electrical Engineering as a joint degree between Politecnico di Torino, Grenoble INP-Phelma and EPFL Lausanne, in 2018 and 2020 respectively. %
    \lucaphd{}
    \researchinterests{%
      real-time computing, power management of HPC processors, and energy-efficient processor
      architecture}
\end{IEEEbiography}

\begin{IEEEbiography}[%
    {\includegraphics[width=1in,height=1.25in,clip,keepaspectratio]%
        {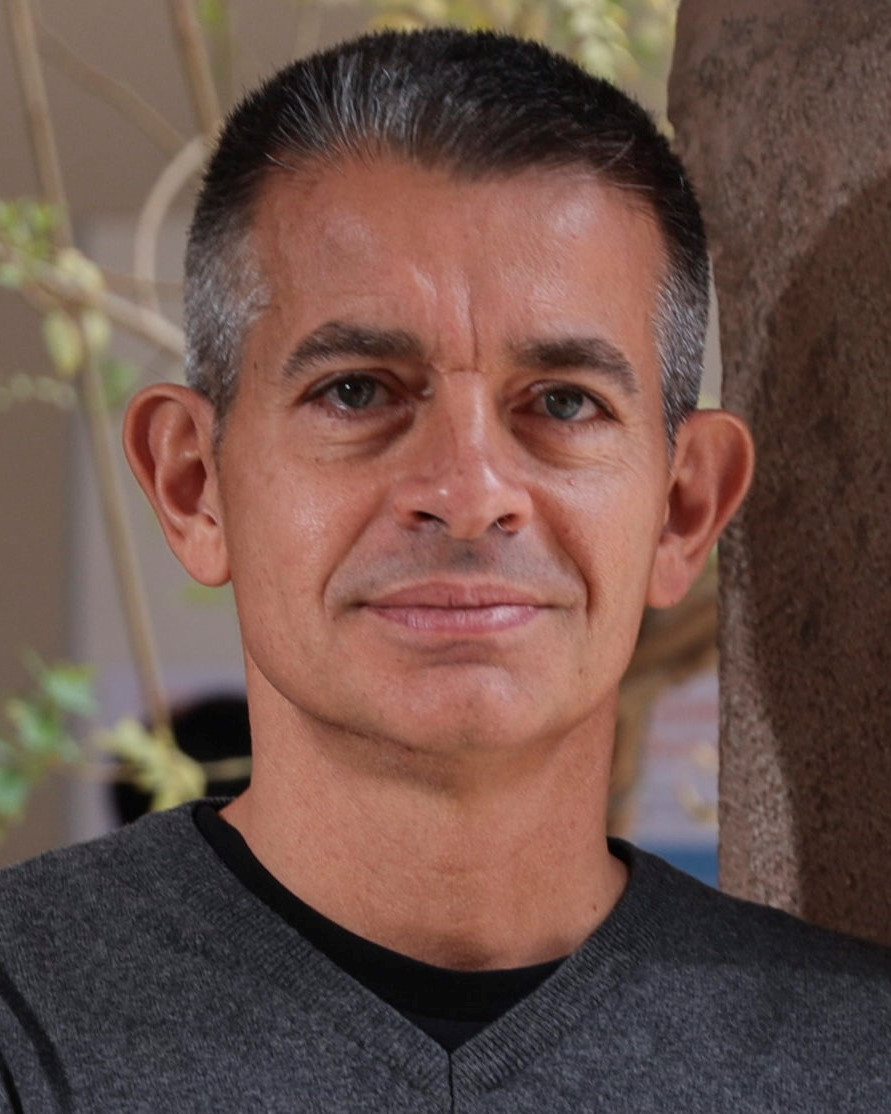}}%
    ]{Luca Benini}
    holds the chair of digital Circuits and systems at ETHZ and is a Full Professor at the Università di Bologna. He received a Ph.D. from Stanford University. He is a Fellow of the ACM and a member of the Academia Europaea. He is the recipient of the 2023 IEEE CS E.J. McCluskey Award.
\end{IEEEbiography}

\vfill

\end{document}

%% file: acronyms.tex
\newacronym{dtm}{DTM}{dynamic thermal management}
\newacronym{stm}{STM}{static thermal management}
\newacronym{hw}{HW}{hardware}
\newacronym{sw}{SW}{software}
\newacronym{ca}{CA}{command/address}
\newacronym{ip}{IP}{intellectual property}
\newacronym{ddr}{DDR}{double data rate}
\newacronym{lpddr}{LPDDR}{low-power double data rate}
\newacronym{rpc}{RPC}{reduced pin count}
\newacronym{dma}{DMA}{direct memory access}
\newacronym{axi}{AXI}{Advanced eXtensible Interface}
\newacronym{dram}{DRAM}{dynamic random access memory}
\newacronym[firstplural=static random access memories (SRAMs)]{sram}{SRAM}{static random access memory}
\newacronym{edram}{eDRAM}{embedded DRAM}
\newacronym[firstplural=systems on chip (SoCs)]{soc}{SoC}{system on chip}
\newacronym{mpsoc}{MPSoC}{multi-processor system on chip}
\newacronym{hesoc}{HeSoC}{heterogeneous system on chip}
\newacronym{sip}{SiP}{system in package}
\newacronym{fpga}{FPGA}{field-programmable gate array}
\newacronym{asic}{ASIC}{application-specific integrated circuit}
\newacronym{phy}{PHY}{physical layer}
\newacronym{ml}{ML}{machine learning}
\newacronym{iot}{IoT}{internet of things}
\newacronym{foss}{FOSS}{free and open source}
\newacronym{cmos}{CMOS}{complementary metal-oxide-semiconductor}
\newacronym{sut}{SUT}{system under test}
\newacronym{isut}{ISUT}{integrated system under test}
\newacronym{rtl}{RTL}{register transfer level}
\newacronym{hil}{HIL}{hardware in the loop}
\newacronym{pil}{PIL}{processor in the loop}
\newacronym{fil}{FIL}{FPGA in the loop}
\newacronym{mil}{MIL}{model in the loop}
\newacronym{sil}{SIL}{software in the loop}
\newacronym{hpc}{HPC}{high performance computing}
\newacronym{mcu}{MCU}{microcontroller unit}
\newacronym{fub}{FUB}{functional unit block}
\newacronym{ecu}{ECU}{electronic control unit}
\newacronym{dcu}{DCU}{domain control unit}
\newacronym{zcu}{ZCU}{zonal control unit}
\newacronym{adas}{ADAS}{advanced driver-assistance system}
\newacronym{fame}{FAME}{FPGA Architecture Model Execution}
\newacronym{pl}{PL}{Programmable Logic}
\newacronym{ps}{PS}{Processing System}
\newacronym{apu}{APU}{Application Processing Unit}
\newacronym{ocm}{OCM}{on-chip memory}
\newacronym{pcs}{PCS}{power controller system}
\newacronym{pcf}{PCF}{power control firmware}
\newacronym{plic}{PLIC}{Platform-Level Interrupt Controller}
\newacronym{aplic}{APLIC}{Advanced Platform-Level Interrupt Controller}
\newacronym{pmca}{PMCA}{programmable many-core accelerator}
\newacronym{bram}{BRAM}{block RAM}
\newacronym{lut}{LUT}{look-up table}
\newacronym{ff}{FF}{flip-flop}
\newacronym{fsbl}{FSBL}{First Stage BootLoader}
\newacronym{pvt}{PVT}{Process, Voltage, Temperature}
\newacronym{hls}{HLS}{high-level synthesis}
\newacronym{mqtt}{MQTT}{Message Queuing Telemetry Transport}
\newacronym{cots}{COTS}{commercial off-the-shelf}
\newacronym{cpu}{CPU}{central processing unit}
\newacronym{gpu}{GPU}{graphic processing unit}
\newacronym{ibmocc}{IBM OCC}{IBM on-chip controller}
\newacronym{clic}{CLIC}{Core-Local Interrupt Controller}
\newacronym{clint}{CLINT}{Core-Local Interruptor}
\newacronym{scmi}{SCMI}{System Control and Management Interface}
\newacronym{os}{OS}{Operating System}
\newacronym{mimo}{MIMO}{multiple-input multiple-output}
\newacronym{bmc}{BMC}{Baseboard Management Controller}
\newacronym{qos}{QoS}{quality of service}
\newacronym{tdp}{TPD}{thermal design power}
\newacronym{dvfs}{DVFS}{dynamic voltage and frequency scaling}
\newacronym{dfs}{DFS}{dynamic frequency scaling}
\newacronym{dvs}{DVS}{dynamic voltage scaling}
\newacronym{rtu}{RTU}{Real Time Unit}
\newacronym{pe}{PE}{processing element}
\newacronym{noc}{NoC}{network on chips}
\newacronym{pid}{PID}{proportional integral derivative}
\newacronym{sota}{SOTA}{state of the art}
\newacronym{fpu}{FPU}{floating point unit}
\newacronym{pcu}{PCU}{Power Control Unit}
\newacronym{scp}{SCP}{System Control Processor}
\newacronym{mcp}{MCP}{Manageability Control Processor}
\newacronym{occ}{OCC}{On-Chip Controller}
\newacronym{smu}{SMU}{System Management Unit}
\newacronym{ap}{AP}{application-class processor}
\newacronym{vrm}{VRM}{voltage regulator module}
\newacronym{pfct}{PFCT}{periodic frequency control task}
\newacronym{pvct}{PVCT}{periodic voltage control task}
\newacronym{ipc}{IPC}{instructions per cycle}
\newacronym{simd}{SIMD}{single instruction, multiple data}
\newacronym{mctp}{MCTP}{Management Component Transport Protocol}
\newacronym{pldm}{PLDM}{Platform Level Data Model}
\newacronym{rtos}{RTOS}{real-time operating system}
\newacronym{gpos}{GPOS}{general-purpose operating system}
\newacronym{hlc}{HLC}{high-level controller}
\newacronym{llc}{LLC}{low-level controller}
\newacronym{isr}{ISR}{interrupt service routine}
\newacronym{wcet}{WCET}{worst-case execution time}
\newacronym{mcs}{MCS}{mixed criticality system}
\newacronym{isa}{ISA}{instruction set architecture}
\newacronym{csr}{CSR}{Control and Status Register}
\newacronym{apcs}{APCS}{Arm procedure call standard}
\newacronym{fsm}{FSM}{finite state machine}
\newacronym{fiq}{FIQ}{fast interrupt request}
\newacronym{irq}{IRQ}{standard interrupt request}
\newacronym{nvic}{NVIC}{nested vectored interrupt controller}
\newacronym{vic}{VIC}{vectored interrupt controller}
\newacronym{gic}{GIC}{generic interrupt controller}
\newacronym{ge}{GE}{gate equivalent}
\newacronym{icu}{ICU}{interrupt control unit}
\newacronym{srn}{SRN}{service request node}
\newacronym{hart}{HART}{hardware thread}
\newacronym{ir}{IR}{interrupt router}
\newacronym{pc}{PC}{program counter}
\newacronym{intid}{INTID}{interrupt identification number}
\newacronym{csa}{CSA}{context save area}

\newacronym [ longplural={scratchpad memories} ] {spm}{SPM}{scratchpad memory}
\newacronym{shv}{SHV}{selective hardware vectoring}
\newacronym{abi}{ABI}{application binary interface}
\newacronym{eabi}{EABI}{embedded application binary interface}
\newacronym{imsic}{IMSIC}{incoming message signaled interrupt controller}
\newacronym{id}{ID}{instruction decode}
\newacronym{if}{IF}{instruction fetch}
\newacronym{ex}{EX}{execution}

%% file: main.bbl
\begin{thebibliography}{10}
\providecommand{\url}[1]{#1}
\csname url@samestyle\endcsname
\providecommand{\newblock}{\relax}
\providecommand{\bibinfo}[2]{#2}
\providecommand{\BIBentrySTDinterwordspacing}{\spaceskip=0pt\relax}
\providecommand{\BIBentryALTinterwordstretchfactor}{4}
\providecommand{\BIBentryALTinterwordspacing}{\spaceskip=\fontdimen2\font plus
\BIBentryALTinterwordstretchfactor\fontdimen3\font minus
  \fontdimen4\font\relax}
\providecommand{\BIBforeignlanguage}[2]{{%
\expandafter\ifx\csname l@#1\endcsname\relax
\typeout{** WARNING: IEEEtran.bst: No hyphenation pattern has been}%
\typeout{** loaded for the language `#1'. Using the pattern for}%
\typeout{** the default language instead.}%
\else
\language=\csname l@#1\endcsname
\fi
#2}}
\providecommand{\BIBdecl}{\relax}
\BIBdecl

\bibitem{rochange_2014}
\BIBentryALTinterwordspacing
C.~Rochange, S.~Uhrig, and P.~Sainrat, \emph{Time-Predictable Architectures},
  ser. {FOCUS} - Computer Engineering Series.\hskip 1em plus 0.5em minus
  0.4em\relax iSTE / Wiley, 2014. [Online]. Available:
  \url{http://eu.wiley.com/WileyCDA/WileyTitle/productCd-1848215932.html}
\BIBentrySTDinterwordspacing

\bibitem{marongiu_survey_2018}
L.~M. Pinho, E.~Quinones, M.~Bertogna, A.~Marongiu, V.~Nelis, P.~Gai, and
  J.~Sancho, \emph{High-Performance and Time-Predictable Embedded
  Computing}.\hskip 1em plus 0.5em minus 0.4em\relax Wharton, TX, USA: River
  Publishers, 2018.

\bibitem{linux-preempt-rt}
\BIBentryALTinterwordspacing
F.~Reghenzani, G.~Massari, and W.~Fornaciari, ``{The Real-Time Linux Kernel: A
  Survey on PREEMPT\_RT},'' \emph{ACM Comput. Surv.}, vol.~52,
  \href{http://dx.doi.org/10.1145/3297714}{DOI 10.1145/3297714}, no.~1, Feb.
  2019. [Online]. Available: \url{https://doi.org/10.1145/3297714}
\BIBentrySTDinterwordspacing

\bibitem{miao_2011}
M.~Liu, D.~Liu, Y.~Wang, M.~Wang, and Z.~Shao, ``On improving real-time
  interrupt latencies of hybrid operating systems with two-level hardware
  interrupts,'' \emph{IEEE Transactions on Computers}, vol.~60,
  \href{http://dx.doi.org/10.1109/TC.2010.119}{DOI 10.1109/TC.2010.119}, no.~7,
  pp. 978--991, 2011.

\bibitem{rtai_2000}
P.~Mantegazza, E.~L. Dozio, and S.~Papacharalambous, ``Rtai: Real time
  application interface,'' \emph{Linux J.}, vol. 2000, no. 72es, p. 10–es,
  Apr. 2000.

\bibitem{ramam_94}
K.~Ramamritham and J.~Stankovic, ``Scheduling algorithms and operating systems
  support for real-time systems,'' \emph{Proc. IEEE}, vol.~82,
  \href{http://dx.doi.org/10.1109/5.259426}{DOI 10.1109/5.259426}, no.~1, pp.
  55--67, 1994.

\bibitem{minimize_int}
\BIBentryALTinterwordspacing
D.~Kleidermacher, ``Minimizing interrupt response time,'' 2005. [Online].
  Available:
  \url{http://web.engr.oregonstate.edu/~traylor/ece473/pdfs/minimize_interrupt_response_time.pdf}
\BIBentrySTDinterwordspacing

\bibitem{intro_embedded}
J.~Valvano, \emph{Introduction to Embedded Systems}, 08 2016.

\bibitem{WCET_REALLOCATION_TVLSI}
Y.~Huang, L.~Shi, J.~Li, Q.~Li, and C.~J. Xue, ``Wcet-aware re-scheduling
  register allocation for real-time embedded systems with clustered vliw
  architecture,'' \emph{IEEE Transactions on Very Large Scale Integration
  (VLSI) Systems}, vol.~22,
  \href{http://dx.doi.org/10.1109/TVLSI.2012.2236114}{DOI
  10.1109/TVLSI.2012.2236114}, no.~1, pp. 168--180, 2014.

\bibitem{xiangrong_2006}
\BIBentryALTinterwordspacing
X.~Zhou and P.~Petrov, ``Rapid and low-cost context-switch through embedded
  processor customization for real-time and control applications,'' in
  \emph{Proceedings of the 43rd Annual Design Automation Conference}, ser. DAC
  '06, \href{http://dx.doi.org/10.1145/1146909.1147001}{DOI
  10.1145/1146909.1147001}, p. 352–357.\hskip 1em plus 0.5em minus
  0.4em\relax New York, NY, USA: Association for Computing Machinery, 2006.
  [Online]. Available: \url{https://doi.org/10.1145/1146909.1147001}
\BIBentrySTDinterwordspacing

\bibitem{behnke_2020}
\BIBentryALTinterwordspacing
I.~Behnke, L.~Pirl, L.~Thamsen, R.~Danicki, A.~Polze, and O.~Kao,
  ``Interrupting real-time {IoT} tasks: How bad can it be to connect your
  critical embedded system to the internet?'' in \emph{2020 {IEEE} 39th
  International Performance Computing and Communications Conference ({IPCCC})},
  \href{http://dx.doi.org/10.1109/ipccc50635.2020.9391536}{DOI
  10.1109/ipccc50635.2020.9391536}.\hskip 1em plus 0.5em minus 0.4em\relax
  {IEEE}, Nov. 2020. [Online]. Available:
  \url{https://doi.org/10.1109%2Fipccc50635.2020.9391536}
\BIBentrySTDinterwordspacing

\bibitem{AUTOMOTIVE_DYNAMIQ_MPAM}
F.~Rehm, J.~Seitter, J.-P. Larsson, S.~Saidi, G.~Stea, R.~Zippo, D.~Ziegenbein,
  M.~Andreozzi, and A.~Hamann, ``The road towards predictable automotive high -
  performance platforms,'' in \emph{2021 Design, Automation \& Test in Europe
  Conference \& Exhibition (DATE)},
  \href{http://dx.doi.org/10.23919/DATE51398.2021.9473996}{DOI
  10.23919/DATE51398.2021.9473996}, pp. 1915--1924, 2021.

\bibitem{asanovic_riscv}
\BIBentryALTinterwordspacing
K.~Asanović and D.~A. Patterson, ``Instruction sets should be free: The case
  for risc-v,'' EECS Department, University of California, Berkeley, Tech. Rep.
  UCB/EECS-2014-146, Aug. 2014. [Online]. Available:
  \url{http://www2.eecs.berkeley.edu/Pubs/TechRpts/2014/EECS-2014-146.html}
\BIBentrySTDinterwordspacing

\bibitem{RISCV_II}
\BIBentryALTinterwordspacing
A.~Waterman, Y.~Lee, R.~Avizienis, D.~A. Patterson, and K.~Asanović, ``The
  risc-v instruction set manual volume ii: Privileged architecture version
  1.9,'' EECS Department, University of California, Berkeley, Tech. Rep.
  UCB/EECS-2016-129, Jul. 2016. [Online]. Available:
  \url{http://www2.eecs.berkeley.edu/Pubs/TechRpts/2016/EECS-2016-129.html}
\BIBentrySTDinterwordspacing

\bibitem{clic}
\BIBentryALTinterwordspacing
RISC-V, ``"smclic" core-local interrupt controller (clic) risc-v privileged
  architecture extension.'' [Online]. Available:
  \url{https://github.com/riscv/riscv-fast-interrupt/blob/master/clic.adoc}
\BIBentrySTDinterwordspacing

\bibitem{RISCY_CORE_TVLSI}
M.~Gautschi, P.~D. Schiavone, A.~Traber, I.~Loi, A.~Pullini, D.~Rossi,
  E.~Flamand, F.~K. Gürkaynak, and L.~Benini, ``Near-threshold risc-v core
  with dsp extensions for scalable iot endpoint devices,'' \emph{IEEE
  Transactions on Very Large Scale Integration (VLSI) Systems}, vol.~25,
  \href{http://dx.doi.org/10.1109/TVLSI.2017.2654506}{DOI
  10.1109/TVLSI.2017.2654506}, no.~10, pp. 2700--2713, 2017.

\bibitem{CV32E40P_manual}
\BIBentryALTinterwordspacing
D.~Schiavone, \emph{OpenHW Group CV32E40P User Manual}, OpenHW Group. [Online].
  Available: \url{https://cv32e40p.readthedocs.io/en/latest/}
\BIBentrySTDinterwordspacing

\bibitem{nuclei_isa}
\BIBentryALTinterwordspacing
N.~S. Technology, \emph{Nuclei ISA spec}, Nuclei System Technology, 2021.
  [Online]. Available:
  \url{https://doc.nucleisys.com/nuclei_spec/isa/introduction.html}
\BIBentrySTDinterwordspacing

\bibitem{Ottaviano2023ControlPULPAR}
A.~Ottaviano, R.~Balas, G.~Bambini, A.~del Vecchio, M.~Ciani, D.~Rossi,
  L.~Benini, and A.~Bartolini, ``Controlpulp: A risc-v on-chip parallel power
  controller for many-core hpc processors with fpga-based hardware-in-the-loop
  power and thermal emulation,'' \emph{ArXiv}, vol. abs/2306.09501, 2023.

\bibitem{freertos}
\BIBentryALTinterwordspacing
R.~Barry. Freertos: Real-time operating system for microcontrollers. Real Time
  Engineers Ltd. (2023). [Online]. Available:
  \url{https://www.freertos.org/index.html}
\BIBentrySTDinterwordspacing

\bibitem{lin_10}
C.-M. Lin, ``Nested interrupt analysis of low cost and high performance
  embedded systems using gspn framework,'' \emph{IEICE Transactions}, vol.
  93-D, \href{http://dx.doi.org/10.1587/transinf.E93.D.2509}{DOI
  10.1587/transinf.E93.D.2509}, pp. 2509--2519, 09 2010.

\bibitem{riscv_plic}
\BIBentryALTinterwordspacing
R.-V.~T. Group, \emph{RISC-V Platform-Level Interrupt Controller
  Specification}, RISC-V International, 2023. [Online]. Available:
  \url{https://github.com/riscv/riscv-plic-spec/blob/master/riscv-plic-1.0.0.pdf}
\BIBentrySTDinterwordspacing

\bibitem{yiu_cortex_reference_13}
J.~Yiu, \emph{The Definitive Guide to ARM Cortex-M3 and Cortex-M4 Processors,
  Third Edition}, 3rd~ed.\hskip 1em plus 0.5em minus 0.4em\relax USA: Newnes,
  2013.

\bibitem{cortex_m4_manual}
\BIBentryALTinterwordspacing
Arm, \emph{Cortex-M4 Technical Reference Manual}, Arm, 2020. [Online].
  Available: \url{https://developer.arm.com/documentation/100166/0001/}
\BIBentrySTDinterwordspacing

\bibitem{arm_interrupt_beginner_guide}
\BIBentryALTinterwordspacing
J.~Yiu, \emph{A Beginner’s Guide on Interrupt Latency - and Interrupt Latency
  of the Arm Cortex-M processors}, Arm, 2012. [Online]. Available:
  \url{https://community.arm.com/arm-community-blogs/b/architectures-and-processors-blog/posts/beginner-guide-on-interrupt-latency-and-interrupt-latency-of-the-arm-cortex-m-processors?pifragment-22714=2}
\BIBentrySTDinterwordspacing

\bibitem{TASK_SWITCHING_TVLSI}
V.~G. Gaitan, N.~C. Gaitan, and I.~Ungurean, ``Cpu architecture based on a
  hardware scheduler and independent pipeline registers,'' \emph{IEEE
  Transactions on Very Large Scale Integration (VLSI) Systems}, vol.~23,
  \href{http://dx.doi.org/10.1109/TVLSI.2014.2346542}{DOI
  10.1109/TVLSI.2014.2346542}, no.~9, pp. 1661--1674, 2015.

\bibitem{mao_2021}
B.~Mao, N.~Tan, T.~Chong, and L.~Li, ``A clic extension based fast interrupt
  system for embedded risc-v processors,'' in \emph{2021 6th International
  Conference on Integrated Circuits and Microsystems (ICICM)},
  \href{http://dx.doi.org/10.1109/ICICM54364.2021.9660345}{DOI
  10.1109/ICICM54364.2021.9660345}, pp. 109--113, 2021.

\bibitem{balas_2021}
R.~Balas and L.~Benini, ``Risc-v for real-time mcus - software optimization and
  microarchitectural gap analysis,'' in \emph{2021 Design, Automation, Test in
  Europe Conference Exhibition (DATE)},
  \href{http://dx.doi.org/10.23919/DATE51398.2021.9474114}{DOI
  10.23919/DATE51398.2021.9474114}, pp. 874--877, 2021.

\bibitem{renesas_m32c80}
\BIBentryALTinterwordspacing
Renesas, \emph{Hardware Manual - RENESAS MCU M16C FAMILY / M32C/80 SERIES},
  Renesas, 2023. [Online]. Available:
  \url{https://www.renesas.com/us/en/document/mah/m32c87-group-m32c87-m32c87a-m32c87b-hardware-manual}
\BIBentrySTDinterwordspacing

\bibitem{sifive_e21}
\BIBentryALTinterwordspacing
S.~Inc., \emph{SiFive E21 Core Complex Manual}, SiFive Inc., 2021. [Online].
  Available:
  \url{https://sifive.cdn.prismic.io/sifive/7c22c2ec-8af4-4b6c-a5fe-9327d91e7808_e21_core_complex_manual_21G1.pdf}
\BIBentrySTDinterwordspacing

\bibitem{stm32l5}
\BIBentryALTinterwordspacing
STMicroelectronics, \emph{STM32L5 - NVIC}, STMicroelectronics, 2020. [Online].
  Available:
  \url{https://www.st.com/content/ccc/resource/training/technical/product_training/group1/61/35/d2/07/34/6f/4e/83/STM32L5-System-Nested_Vectored_Interrupt_Control_NVIC/files/STM32L5-System-Nested_Vectored_Interrupt_Control_NVIC.pdf/_jcr_content/translations/en.STM32L5-System-Nested_Vectored_Interrupt_Control_NVIC.pdf}
\BIBentrySTDinterwordspacing

\bibitem{tricore_hotchips16}
\BIBentryALTinterwordspacing
H.~C.~. Infineon Technologies~AG, ``A fast powertrain microcontroller.''
  [Online]. Available:
  \url{https://old.hotchips.org/wp-content/uploads/hc_archives/hc16/3_Tue/6_HC16_Sess7_Pres1_bw.pdf}
\BIBentrySTDinterwordspacing

\bibitem{aurix_tc27x}
\BIBentryALTinterwordspacing
I.~T. AG, \emph{TC27x D-Step, 32-Bit Single-Chip Microcontroller}, Infineon
  Technologies AG, 2014. [Online]. Available:
  \url{https://hitex.co.uk/fileadmin/uk-files/downloads/ShieldBuddy/tc27xD_um_v2.2.pdf}
\BIBentrySTDinterwordspacing

\bibitem{aurix_tricore_v1.6}
\BIBentryALTinterwordspacing
I.~T. AG, \emph{TriCore V1.6, Core architecture}, Infineon Technologies AG,
  2012. [Online]. Available:
  \url{https://www.infineon.com/dgdl/tc1_6__architecture_vol1.pdf?fileId=db3a3043372d5cc801373b0f374d5d67}
\BIBentrySTDinterwordspacing

\bibitem{tricore_patent}
I.~T. AG, ``Task context switching rtos,'' U.S. Patent US7434222B2, Oct. 2008.

\bibitem{eclic}
\BIBentryALTinterwordspacing
N.~S. T.~C. Ltd., \emph{ECLIC Unit Introduction}, Nuclei System Technology Co.
  Ltd. [Online]. Available:
  \url{https://doc.nucleisys.com/nuclei_spec/isa/eclic.html}
\BIBentrySTDinterwordspacing

\end{thebibliography}
